\documentclass[fleqn,usenatbib,useAMS]{mnras}

\usepackage{newtxtext,newtxmath}

\usepackage[T1]{fontenc}
\usepackage{ae,aecompl}

\usepackage{graphicx}	
\usepackage{subfig}
\usepackage{amsmath}	
\usepackage{amssymb}	

\usepackage{color, soul}

\newcommand{\e}{\mathrm{e}}
\newcommand{\dd}{\mathrm{d}}

\newcommand{\eps}{\mathrm{\epsilon}}

\defcitealias{Georgiou2019b}{G19}
\defcitealias{Johnston2019}{J19}
\defcitealias{SchneiderBridle2010}{SB10}

\title[IA halo model for cosmic shear analyses]{The halo model as a versatile tool to predict intrinsic alignments}

\author[M.C. Fortuna et al.]{
Maria Cristina Fortuna$^{1}$\thanks{E-mail: fortuna@strw.leidenuniv.nl (Leiden Observatory)},
Henk Hoekstra$^{1}$,
Benjamin Joachimi$^{2}$,
Harry Johnston$^{2}$,\newauthor
Nora Elisa Chisari$^{3}$,
Christos Georgiou$^{1}$,
Constance Mahony$^{2}$
\\
$^{1}$Leiden Observatory, Leiden University, PO Box 9513, Leiden, NL-2300 RA, The Netherlands\\
$^{2}$Department of Physics and Astronomy, University College London, Gower Street, London WC1E 6BT, UK\\
$^{3}$Institute for Theoretical Physics, Utrecht University, Princetonplein 5, 3584 CE Utrecht,
The Netherlands\\
\newline
}

\date{Accepted XXX. Received YYY; in original form ZZZ}

\pubyear{2020}

\begin{document}
\label{firstpage}
\pagerange{\pageref{firstpage}--\pageref{lastpage}}
\maketitle

\begin{abstract}
Intrinsic alignments (IAs) of galaxies are an important contaminant for cosmic shear studies, but the modelling is complicated by the dependence of the signal on the source galaxy sample. In this paper, we use the halo model formalism to capture this diversity and examine its implications for Stage-III and Stage-IV cosmic shear surveys. We account for the different IA signatures at large and small scales, as well for the different contributions from central/satellite and red/blue galaxies, and we use realistic mocks to account for the characteristics of the galaxy populations as a function of redshift. We inform our model using the most recent observational findings: we include a luminosity dependence at both large and small scales and a radial dependence of the signal within the halo. We predict the impact of the total IA signal on the lensing angular power spectra, including the current uncertainties from the IA best-fits to illustrate the range of possible impact on the lensing signal: the lack of constraints for fainter galaxies is the main source of uncertainty for our predictions of the IA signal. We investigate how well effective models with limited degrees of freedom can account for the complexity of the IA signal. Although these lead to negligible biases for Stage-III surveys, we find that, for Stage-IV surveys, it is essential to at least include an additional parameter to capture the redshift dependence.

\end{abstract}





\section{Introduction}

As the light of distant galaxies travels towards us, it is deflected by matter inhomogeneities. The cumulative effect of these small distortions leads to a preferential apparent alignment of galaxy shapes, a phenomenon called weak lensing. The resulting correlation of galaxy shapes (\emph{cosmic shear}) provides direct information on the matter distribution in the Universe as well as the effect of dark energy on the geometry and the growth of structures \citep[e.g][]{Bartelmann&Schneider2001, Kibinger2015review}. However, extracting cosmological parameter estimates from weak lensing surveys is challenging due to a number of systematic errors it is prone to.  
On the measurements side, the main sources of bias come from the uncertainty in the source redshift distributions and the actual shape measurements, for which great improvements have been achieved in the last decades, due to advances in both image simulations and shape measurement algorithms \citep[][for a dedicated review]{Kannawadi2019, Mandelbaum2018review}.

On the modelling side, a naive interpretation of cosmic shear would relate the observed correlations between galaxy orientations as solely arising from the lensing effect of matter. In reality, galaxies form and live inside dark matter haloes and they are continuously exposed to the gravitational interaction with the surrounding matter distribution. This leads to the coherent alignment induced by the underlying tidal field on physically near galaxies, the so-named intrinsic alignment (IA) \citep[][for extensive reviews]{Joachimi2015review, Kiessling2015review, Kirk2015review, Troxel2015review}. If not properly accounted, IA can affect the inferred properties of the matter distribution from lensing. In the perspective of high precision surveys such as Euclid\footnote{\url{https://www.euclid-ec.org}} \citep{Laureijs2011} and LSST\footnote{\url{https://www.lsst.org}} \citep{Abell2009lsst}, which aim to measure cosmological parameters with an accuracy better than a percent, it is crucial to properly model the impact of IA and to quantify the level of precision required in our models and IA constraints \citep{ JoachimiBridle2010J, Kirk2010, Krause2016}.

One of the challenges in mitigating the effect of IA comes from the differences between the samples employed in studies of IA and in cosmic shear. Pressure supported (red/elliptical) galaxies are more subjected to the effect of tidal fields and tend to stretch their shapes in the direction of the matter overdensities \citep{Catelan2001}. This turns into a non-negligible IA signal, observationally constrained by a number of works \citep[e.g.][hereafter \citetalias{Johnston2019}]{Mandelbaum2006a, Hirata2007, Okumura2009, Okumura&Jing2009, Joachimi2011b, Singh2015, Johnston2019}. On the other hand, disc, rotationally supported (blue) galaxies preferentially align their spins through a torque mechanism. Although this has been observed in simulations, there is no consensus on the final predictions due to the different implementations of hydrodynamics and baryonic feedback \citep[e.g.][]{Chisari2015, Tenneti2016, Codis2018, Kralijc2020}. From an observational point of view, the alignment of blue galaxies has not been detected yet, neither at low and intermediate redshifts \citep[][\citetalias{Johnston2019}]{Mandelbaum2011WiggleZ, Blazek2012GGLIA, Samuroff2018DES}, nor at high redshifts \citep{Tonegawa2018}. 

For this reason, to maximise the signal-to-noise ratio, the majority of IA analyses focus on low redshift red galaxies, while cosmic shear surveys typically span a much broader range in redshift and do not make any colour selection. A proper re-scaling of IA predictions, weighted by the fraction of red galaxies in the sample, is then required in order to correctly account for the alignment contribution to the signal. 

While the aforementioned alignment mechanisms describe the behaviour of the central galaxies well, the picture at small scales is complicated by the intra-halo tidal fields, galaxy mergers and halo assembly history, as well as AGN feedback and winds \citep{Soussana2020, Tenneti2017}. \citet{Pereira2008} and \citet{Pereira2010} investigated the satellite halo alignment in simulations, finding an overall tendency of satellites to point radially towards the centre of the host halo, due to a continuous torquing mechanism that aligns their major axes in the direction of the gravitational potential gradient during their orbits. Motivated by their findings, a halo model description of this alignment term was developed by \citet[][hereafter \citetalias{SchneiderBridle2010}]{SchneiderBridle2010}. However, \citet{Sifon2015} did not find observational evidence for satellite alignment in clusters. Similarly, \citet{Chisari2014} explored the alignment signal around stacked clusters and found it to be consistent with zero. \citet{Huang2018} pointed out that the signal depends on the shape algorithm used, a feature further confirmed by \citet[][hereafter \citetalias{Georgiou2019b}]{Georgiou2019b}.

Recently, \citetalias{Johnston2019} and \citetalias{Georgiou2019b} investigated the alignment signal in the overlapping region between the Kilo Degree Survey\footnote{\url{http://kids.strw.leidenuniv.nl}} \citep[KiDS,][]{deJong2013, Kuijken2019DR4} and the Galaxy and Mass Assembly survey \citep[GAMA,][]{Driver2011}. The detected IA signal provides evidence that a simple dichotomy between red and blue galaxies is not sufficient to capture the entire physics of the IA signal. In particular, \citetalias{Georgiou2019b} observed a scale dependence of the satellite alignment, with satellite shapes radially aligned at small radii and a vanishing signal towards larger scales. As a consequence, the tendency of satellites to be randomly orientated at large scales suppresses the the overall IA signal, as observed by \citetalias{Johnston2019}. Therefore, even in the linear regime, where the IA signal can be modelled through the linear alignment model \citep[LA,][]{Hirata2004}, the evolution of the satellite fraction in the sample can imprint a varying amplitude to the signal, a feature never explored by any forecasting analysis so far. 

Cosmic shear analyses employ tomographic binning to investigate the growth of structures and better constrain cosmological parameters. Since these surveys are flux-limited, the tomography imprints an indirect galaxy selection, including only the most luminous galaxies in the high-redshift bins. As satellites are intrinsically fainter, this turns into a satellite cut at high redshifts. Satellites contribute predominantly a random signal at large scales and therefore can induce a modulation of the IA signal over the bins, suppressing it at large scales and boosting it at small scales, at low redshifts. Similarly, the fraction of red galaxies varies with redshift. The extrapolation of the results from IA studies then requires some care, since the majority of them limit their analyses to low-to-intermediate redshifts.

In addition, a luminosity dependence of the IA signal is currently under debate. While it has been observed for large luminous galaxies \citep{Mandelbaum2006a, Hirata2007, Joachimi2011b, Singh2015}, \citetalias{Johnston2019} has found no evidence for any luminosity scaling, hinting towards a more complex sample dependence. Similarly, two studies suggest a different behaviour for the satellite alignment signal, with \citet{Huang2018} detecting a more prominent alignment for the brightest satellites located close to the central galaxy, while \citetalias{Georgiou2019b} do not observe any luminosity trend in galaxy groups, but confirm a radial dependent signal. As for the large scales, a luminosity dependence of the satellite alignment can significantly change the contamination for a lensing survey, where the low redshift tomographic bins are dominated by faint satellites.

Understanding the sample dependence in the IA mechanism is a key feature to properly model it in the broader case of a cosmic shear galaxy sample. In this paper we investigate the impact of satellite galaxy alignment both at large and small scales. We provide a unified framework to incorporate all of the sample dependencies that emerged from observations, through the halo model formalism. We also explore the areas of tension between different measurements in the literature, trying to incorporate all of the available information as well as the current uncertainties in our predictions. We base our model on \citetalias{SchneiderBridle2010}, including the scale dependent signal measured in \citetalias{Georgiou2019b} and the luminosity dependence suggested by \citet{Huang2018}. 

The paper is organised as following. In Sect. \ref{subsec:MICE} we describe the mock data we use to simulate a cosmic shear survey, for which we employ the Marenostrum Institut de Ci$\grave{\mathrm{e}}$ncies de l'Espai Simulations (MICE). We build our mock to resemble a Stage III survey, mainly inspired by the final data release of KiDS. Sect.~\ref{sec:modelling_red_central_fraction_impact} introduces our model at large scales. We explore the possibility that part of the tension around the luminosity scaling of the IA signal is caused by neglecting the satellite fraction in the samples while modelling the signal. We provide a model that accounts for both the role of satellites and the differences between different data sets: we investigate the compatibility of the measurements in the literature within this framework. In Sect.~\ref{sec:the_impact_of_satellites_at_small_scales} we address the behaviour of satellites at small scales. We re-analyse the \citetalias{Georgiou2019b} measurement in the context of a red/blue distinction of the galaxy population, and model the satellite alignment including both a radial and luminosity dependence. In Sect.~\ref{sec:results} we show the predicted IA signal and illustrate the impact on cosmic shear studies. We investigate the impact of adopting simplistic IA models when performing cosmological analysis and address the level of bias expected for a Stage-III (current generation) and a Stage-IV (next generation) surveys. In Sect.~\ref{sec:conclusions} we draw our conclusions. 

Throughout this paper we assume the MICE cosmology as our cosmological model of reference: a spatially flat $\Lambda$CDM model with $h = 0.7$, $\Omega_m=0.25$, $\Omega_{b} = 0.044$, $\Omega_{\Lambda}=0.75$, $n_s=0.95$, $\sigma_8=0.8$. We use $\bar{\rho}_m$ as the present day mean matter density of the Universe. We provide our predictions and measurements in units of $h$. Absolute magnitudes are always given assuming $h=1$.

\section{MICE simulation}\label{subsec:MICE}

To investigate the impact of red and satellite fractions on the IA signal, we need a realistic representation of the galaxy sample that populates a cosmic shear survey. \citet[][]{Krause2016} has shown that one of the major sources of uncertainties in forecasting IA for future cosmic shear surveys comes from the uncertainty in the luminosity function modelling, which determines the red/blue fraction of galaxies in the analysis. In this work, we make use of the MICECATv2.0 \footnote{MICECAT v2 is publicly available at \url{https://cosmohub.pic.es/home}} simulation \citep{Fosalba2015b} as a realisation of our Universe and select galaxies based on the typical values of redshift, magnitude and area for a Stage III survey. We use the simulations as our reference cosmic shear survey, for which we can extract all the necessary information.

MICECAT is a public catalogue, now at its second data release, created to reproduce a number of local observational constraints and it is for this reason particularly suitable for our purposes. The mock galaxy catalogue is obtained from an N-body simulation containing $7\times10^{10}$ dark matter particles in a $(3072 h^{-1} \mathrm{Mpc})^3$ comoving volume \citep{Fosalba2015b} and then populated using a hybrid implementation of Halo Occupation Distribution (HOD) and Sub-Halo Abundance Matching (SHAM) \citep[]{Crocce2015, Carretero2015}. 

Given the importance of having robust satellite fractions per luminosity and redshift bin and a representative colour distribution for our analysis, we report here the most relevant features adopted in \citet{Carretero2015} to build the galaxy catalogue. The HOD parametrisation employed to populate the haloes is inspired by \citet{Zheng2005}, with some modifications that we briefly describe here. The HOD provides the probability $P(N_\mathrm{g}|M_\mathrm{h})$ that a halo of a given mass $M_\mathrm{h}$ contains $N_\mathrm{g}$ galaxies of a certain type (central, satellite). To assign galaxies to a halo, a sharp mass-threshold is adopted, such that every halo more massive than $M_\mathrm{min}$ contains at least one (central) galaxy. The number of satellite galaxies follows a Poisson distribution with mean $\langle N_\mathrm{sat} \rangle = \left[ M_\mathrm{h}/M_1 \right]^{\alpha}$. The slope of the power law is chosen to be $\alpha=1$, as constrained by observations \citep[e.g][]{Kravtsov2004, Zehavi2011}, while the mass threshold for satellite galaxies, $M_1$, is modelled to be a function of $M_\mathrm{min}$ and the halo mass $M_h$. The parameters of the functions that relate $M_1$ to $M_\mathrm{min}$ are those that best reproduce the observed galaxy clustering as a function of luminosity in the SDSS \citep{Zehavi2011}. Galaxy luminosities are assigned using abundance matching, based on the observed luminosity function from \citet{Blanton2003} and \citet{Blanton2005} for the faint end. Note that by construction, satellite galaxies are forced to be fainter than 1.05 times the luminosity of their central galaxy. Colours are assigned following an approach similar to \citet{Skibba&Sheth2009}: the colour-magnitude diagram is parametrised using three Gaussians, corresponding to the red, green and blue population; the mean and standard deviations vary as a function of luminosity. The colour of a galaxy is then drawn from these distributions, taking into account its type (satellite or central). The colour-assignment process is calibrated to reproduce the clustering as a function of colour and luminosity in the SDSS \citep{Zehavi2011}. In our analysis we combine the green and blue population, isolating the red sequence with a different cut than what is reported in \citet{Carretero2015}, as discussed in Section \ref{subsec:galaxy_mocks}.

The second release of MICE increases the luminosity range by populating halos/groups with a fewer number of particles with respect to the v1, up to haloes composed by only two particles. Although the abundance of these small groups is not representative of the abundance of haloes at the equivalent halo mass, this is then corrected by the abundance matching.

\subsection{Galaxy mocks}\label{subsec:galaxy_mocks}

We generate two galaxy mocks: the first one reproduces a generic Stage III survey and is employed as our fiducial cosmic shear-like galaxy distribution. A second mock is constructed for a comparison to the results of \citetalias{Johnston2019} and \citetalias{Georgiou2019b} in KiDSxGAMA, and therefore is designed to reproduce the KiDSxGAMA galaxies used in their analysis. We use that mock to understand and interpret our results at small scales, where we use the measurements as input for our cosmic shear analysis. If not specified otherwise, we always refer to the Stage III mock in this work.

\begin{figure*}
\centering
\subfloat[]{{
\includegraphics[width=0.9\columnwidth]{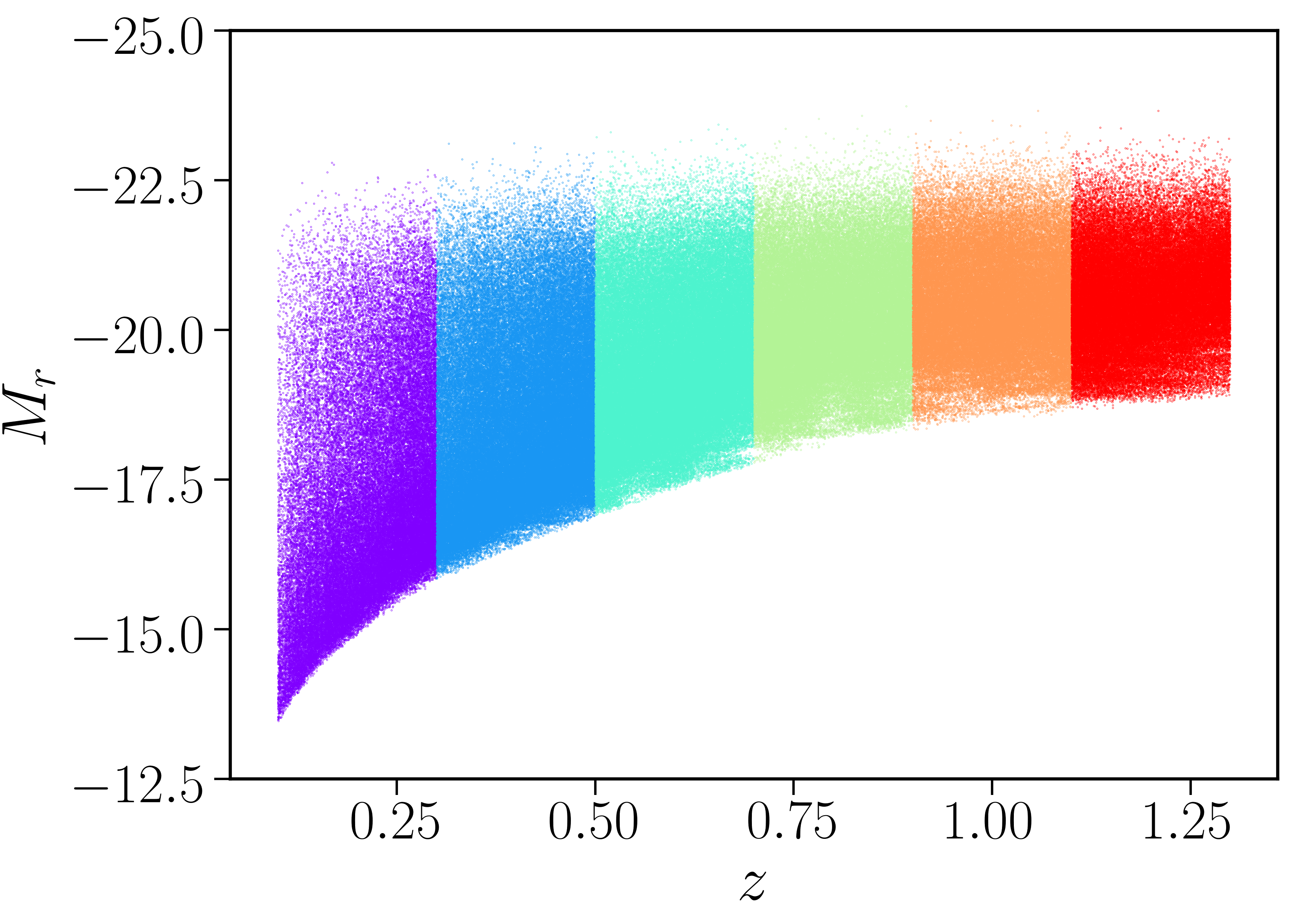}
}}
\qquad
\subfloat[]{{
\includegraphics[width=0.9\columnwidth]{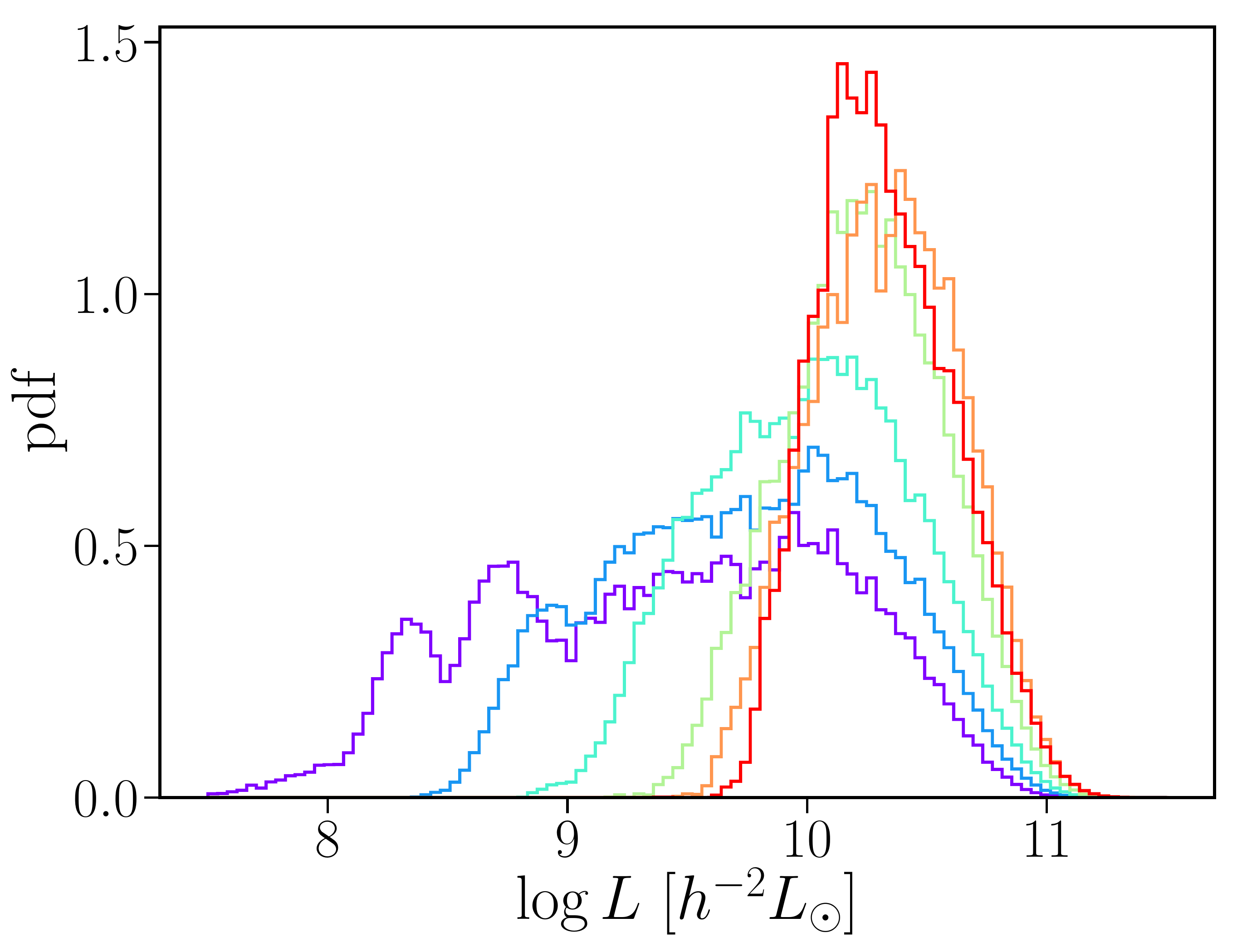}}}

\caption{(a): The distribution in redshift and magnitudes of the sample we select from MICE, with the imposed cut in apparent magnitude at $r<24$. The figure illustrates the samples used in our analysis for the six redshift bins listed in Table \ref{tab:zbin_galinfo}. The plot shows a random selection of $1\%$ of the galaxies in the catalogue. (b): The luminosity distribution of the red central galaxy samples for the six redshift bins, colour coded as in (a).}  
\label{fig:lum_pdf}
\end{figure*}

We select 58 485 848 galaxies from MICECAT v2, covering an area of $1049 \ \mathrm{deg}^2$ in a redshift range $0.1<z<1.3$. We impose a magnitude cut in the SDSS $r-$band $r<24$. We split the sample into six redshift bins with $\Delta z = 0.2$, as shown in Fig. \ref{fig:lum_pdf}a. We correct the magnitudes to take into account the passive evolution of galaxies, as recommended in the MICE readme.

As discussed in the previous section, MICE is complete down to $M_r-5 \log(h) \sim -14$. With our selection, we are close to this limit (the faintest galaxy in our catalogue has a magnitude of $M_r - 5 \log(h) = -13.4$). However, since IA is mainly affected by red galaxies, which are typically brighter than this value, even a small incompleteness should not significantly impact our results.

MICE provides colours in the filters $^{0.1}r$ and $^{0.1}g$. To be consistent with the data-set we aim to compare the mocks to, namely GAMA and the SDSS Main sample used by \citetalias{Johnston2019} and \citetalias{Georgiou2019b}, which select their galaxies using the $g-r$ colour at $z=0$, we correct the MICE $g-r$ colour to be at $z=0$ (F.J. Castander, private communication). We verify that by selecting GAMA-like and SDSS-like mock galaxies from the MICE simulation, imposing the same area coverage, redshift ranges and flux-limit cut, we can reproduce the redshift and magnitude distribution of the samples, the colour-magnitude diagram and the relative galaxy fractions (Appendix \ref{A:satellite_galaxy_fractions_in_mice}). We therefore conclude that MICE galaxies provide a realistic mock for our analysis.  

\begin{figure}
\centering
\includegraphics[width=\columnwidth]{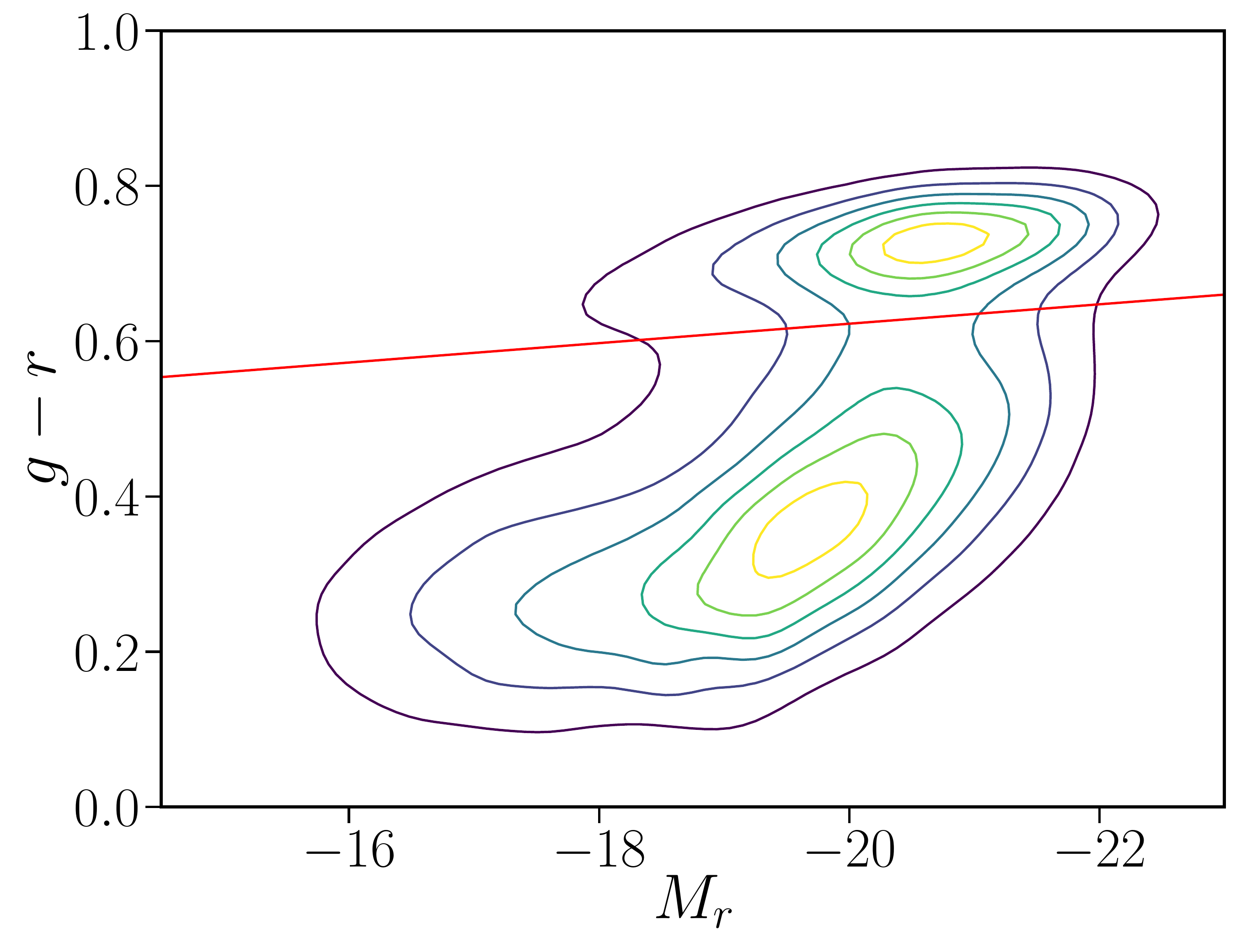}
\caption{The colour-magnitude distribution of the sample. The red line shows the cut at $g-r > 0.61 - 0.0125(M_r + 19)$ we employ to isolate the red sequence. }
\label{fig:redcut}
\end{figure}

To select red galaxies we apply a cut at $g-r > 0.61 - 0.0125(M_r + 19)$, as shown in Fig. \ref{fig:redcut}. The cut qualitatively reproduces the choice in \citetalias{Johnston2019}. Table \ref{tab:zbin_galinfo} summarises the characteristics of our cosmic shear-like galaxy sample in each redshift bin. Due to the flux limit imposed on our sample, the fraction of satellite galaxies drops from low to high redshifts. The red fraction increases for the first three bins, since the faint population is dominated by blue satellites, while it decreases for the last two bins, due to the overall increase of blue galaxies at higher redshifts.

\begin{table*}
	\caption{Properties of the five tomographic bins used in our analysis: the redshift range of each bin ($z_{\mathrm{min}}, z_{\mathrm{max}}$), the number of galaxies ($N_\mathrm{gal}$), the mean luminosity of the red central galaxies in terms of a fiducial luminosity $L_0$ ($ \langle L^\mathrm{red}_\mathrm{cen}\rangle /L_0$), the fraction of satellites in the given bin ($f_\mathrm{sat}$) and the fraction of red galaxies ($f_{\mathrm{red}}$), selected as shown in figure \ref{fig:redcut}. The fiducial luminosity $L_0$ is chosen to be the luminosity corresponding to $M_r=-22$.}
	\label{tab:zbin_galinfo}
	\begin{tabular}{lcccccccr} 
		\hline
		Bin & $z_{\mathrm{min}}$ & $z_{\mathrm{max}}$ & $N_{\mathrm{gal}}$ & $ \langle L^\mathrm{red}_\mathrm{cen} \rangle/L_0$ & $f_{\mathrm{sat}}$ & $f^\mathrm{red}$ & $f^\mathrm{red}_\mathrm{sat}$ & $f^\mathrm{blue}_\mathrm{sat}$\\
		\hline
1 & 0.10 & 0.30 & 7 633 382 & 0.17 & 0.41 & 0.15 & 0.10 & 0.31 \\
2 & 0.30 & 0.50 & 12 445 504 & 0.24 & 0.37 & 0.22 & 0.14 & 0.23 \\
3 & 0.50 & 0.70 & 12 453 204 & 0.33 & 0.33 & 0.27 & 0.16 & 0.17 \\
4 & 0.70 & 0.90 & 9 863 462 & 0.48 & 0.28 & 0.28 & 0.15 & 0.13 \\
5 & 0.90 & 1.10 & 8 003 975 & 0.58 & 0.23 & 0.25 & 0.12 & 0.11 \\
6 & 1.10 & 1.30 & 8 086 321 & 0.55 & 0.22 & 0.26 & 0.13 & 0.09 \\
		\hline
	\end{tabular}
\end{table*}

\section{The impact of satellites at large scales} \label{sec:modelling_red_central_fraction_impact} 

Intrinsic galaxy alignment generates two types of 2-point statistic observables that are relevant in the context of cosmic shear contamination: the correlation between the shapes of two galaxies (II, where `I' stands for Intrinsic) and the correlation between the gravitational shear induced by the lensing effect of a matter inhomogeneity and the intrinsic shape distorted by the same gravitational source (GI, where `G' stands for Gravitation). The final observable is then given by the sum of the cosmic shear power spectrum (GG), which is the one of interest for cosmological studies, and the IA contributions, II and GI.  

\subsection{The linear alignment model}

It is well established that at large scales elliptical galaxies can be modelled through the linear alignment model \citep{Catelan2001, Hirata2004}, which predicts the shape distortion of a galaxy to be proportional to the strength of the tidal field at the moment of its formation.
In Fourier space, the matter-intrinsic and the intrinsic-intrinsic power spectra can thus be written as
\begin{equation}\label{eq:LA}
    P^{\mathrm{LA}}_{\delta \mathrm{I}}(k,z) = - A_\mathrm{IA} C_1 \rho_{c} \frac{\Omega_m}{D(z)} P_{\delta}^{\mathrm{lin}} \ ,
\end{equation}

\begin{equation}\label{eq:LA}
    P^{\mathrm{LA}}_{\mathrm{II}}(k,z) = \left( A_\mathrm{IA} C_1 \rho_{c} \frac{\Omega_m}{D(z)} \right)^2 P_{\delta}^{\mathrm{lin}} \ ,
\end{equation}
where $C_1$ is a normalisation constant, $\rho_c$ the critical density of the Universe today, $D(z)$ the linear growth factor, normalised to unity at $z=0$, and $P_{\delta}^{\mathrm{lin}}$ the linear matter power spectrum. 
We set $C_1 = 5 \times 10^{-14} M_\odot^{-1} h^{-2} \mathrm{Mpc}^{3} $ based on the IA amplitude measured at low redshifts using SuperCOSMOS \citep{Brown2002SuperCOSMOS}.
The free amplitude $A_\mathrm{IA}$ captures any variation with respect to this reference power spectrum. The SuperCOSMOS norm is a common choice in the literature, making the interpretation of our results and any comparison easier. 

A successful modification of this theory replaces the linear matter power spectrum with the non-linear one \citep{BridleKing2007}, so named non-linear linear alignment (NLA). The reason behind the use of the non-linear power spectrum is to partially capture the nonlinear tidal field and it has been shown to fit the measurements better \citep[for example,][]{Blazek2011, Joachimi2011b, Singh2015, Johnston2019}. Although more sophisticated treatments of the non-linear scales have been developed in recent years \citep{Tonegawa2018, Blazek2019, Vlah2020}, and \citet[][]{Samuroff2018DES} have found hints for quadratic alignments in measurements from the Dark Energy Survey (DES), a proper combination of perturbative approaches with the halo model is beyond the scope of this paper. To capture the IA signal at intermediate scales, we therefore use the NLA model (see also Appendix \ref{A:halo_exclusion} for a discussion on the halo exclusion problem in this context).

\subsection{From observations to models: how satellite galaxies complicate the picture}\label{sec:from_observations_to_models_satellites}

As direct measurements of the correlation between the density field and the shear field ($\delta \mathrm{I}$) are not possible, IA studies typically focus on the correlation between the position of a galaxy and the shape of another one, the so called gI term (where `g' stands for galaxy). At large scales, the galaxy position - shear and the matter - shear power spectra are related by the large scale bias, galaxies being  tracers of the underlying matter distribution. For central galaxies, the relation between gI and $\delta$I is simply given by the linear galaxy bias, such that $P_\mathrm{gI}(k) = b_g \ P_{\delta \mathrm{I}}$

A complication arises when interpreting the gI term in the presence of satellite galaxies. Satellites tend to preferentially lie along the major axis of the central galaxy \citep[][ \citetalias{Johnston2019}, \citetalias{Georgiou2019b}]{Huang2016}, which in turn is a proxy for the halo major axis. This anisotropic distribution of satellite positions boosts the satellite position - central shape ($sc$) correlation not only at small scales, but also in the two-halo regime, i.e. when correlating the shape of a central galaxy with the position of a satellite that belongs to a different halo. 

In the context of contamination to lensing, however, such a boost is not expected to have the same importance. Since cosmic shear analyses only correlate shapes, the spatial segregation of satellites is not sufficient to induce a GI signal (where GI is the projected matter - shear power spectrum, i.e. the one that directly contaminate lensing), as satellites need to be coherently oriented to produce a shape correlation \citep[for a discussion on this in simulations, see for example][]{Chisari2015}.

The impact of the anisotropic distribution of satellites has been explored in simulations by \citet{Samuroff2019}, who found a significant enhancement of the signal at small scales and a constant, redshift independent shift at large scales. They found that for an LSST-like (Stage-IV) survey, in the `pessimistic' case (see their section 5.2), this can lead to a shift in the best constrained parameters $\Delta S_8 = 1.4 \sigma$, $\Delta w = 1.5 \sigma$. The recent results from \citetalias{Georgiou2019b}, however, show that at large scales satellite galaxies are randomly oriented with respect to the brightest galaxy in the group, which can be considered as a proxy for the central galaxy, while within the halo their radial alignment is limited to the innermost galaxies. \citetalias{Johnston2019} also found a similar trend when looking at the projected satellite position - shape correlation and central position - shape correlation (their Fig. 7, right panels), suggesting satellites to only coherently orient their shapes in the intra halo regime. Using the same estimator as \citetalias{Georgiou2019b}, $\langle \eps_+ \rangle$, \citet{Sifon2015} found a radial alignment consistent with zero in clusters. Although every detection depends on the choice of the shape measurement algorithm employed, those results suggest that at large scales satellite alignment is a minor contributor compared to the central galaxy alignment. At small scales the picture might be significantly different: we refer the reader to Section \ref{sec:the_impact_of_satellites_at_small_scales} for a discussion on the contribution of satellites at small scales.

While the relative positions of satellites within the halo have a strong impact on the gI correlation, from the argument above, they are not expected to be important in the correlation between the lensed background galaxies and the intrinsically aligned galaxies in the foreground. In this regard, the anisotropic distribution of satellites within the halo complicates the translation of gI measurements to GI, so care has to be taken when adopting informative priors for IA that come from gI measurements. An analysis of this contamination is outside the scope of this paper, and we leave a full modelling of the gI term that can disentangle the two contributions to a forthcoming paper. Given the argument above, we assume that at first order satellites do not contribute to the IA signal at large scales.

\subsection{A weighted linear alignment model}

We have seen in the previous section that, for the sake of accounting for IA contamination in cosmic shear analyses, the role played by satellites is small at large scales. In this sense, central galaxies provide a more consistent picture as they follow the linear alignment mechanism, while the contribution of satellites is mainly to add noise to the measurements. In this context, we can assume that the majority of the contamination comes from the alignment of red central galaxies, while blue central galaxies are expected to add a minor although still very uncertain contribution. 

Motivated by the need of priors for our signals, we decide here to use the NLA model for both red and blue galaxies. In this way, we can directly link our predictions to observational constraints, for which the NLA model has been used to fit the signal \citepalias{Johnston2019}. In principle, if the linear alignment mechanism is truly responsible for the alignment of blue galaxies, a cross term between red and blue galaxies should arise. However, theory suggests that blue galaxies gain their alignment from a torquing mechanism that aligns their spins \citep[][]{Catelan2001, Crittenden2001}, also known as the Quadratic Alignment Model \citep[][]{Hirata2004}. The lack of observational constraints leaves the question of the driving mechanism of blue galaxy alignment open. Our use of the NLA model for blue galaxies should thus be considered as an effective description. We omit the cross term, and consider this approximate treatment sufficient for the scope of the paper, but note that future studies might need to revisit this assumption. The large scale power spectra thus read:

\begin{equation}\label{eq:LA_fraction_weighted_GI}
        P_{\mathrm{\delta I}}^{2h}(k,z) = f^\mathrm{red}_\mathrm{cen} P_{\mathrm{\delta I}, cc}^{2h, \mathrm{red}}(k,z) + f^\mathrm{blue}_\mathrm{cen} P_{\mathrm{\delta I}, cc}^{2h, \mathrm{blue}}(k,z) \ ,
\end{equation}
\begin{equation}\label{eq:LA_fraction_weighted_II}
        P_{\mathrm{II}}^{2h}(k,z) = (f^\mathrm{red}_\mathrm{cen})^2 P_{\mathrm{II}, cc}^{2h, \mathrm{red}}(k,z) + (f^\mathrm{blue}_\mathrm{cen})^2 P_{\mathrm{II}, cc}^{2h, \mathrm{blue}}(k,z) \ ,
\end{equation}
where we have introduced the superscript $2h$ to indicate that these power spectra describe the alignment in the two-halo regime, i.e. for galaxies that do not belong to the same halo (large scale alignment). Similarly, the subscript $cc$ indicates that the correlation only involves central galaxies. $f^\mathrm{red/ blue}_\mathrm{cen}$ is the fraction of red/blue central galaxies in the sample, and we have $f^{\mathrm{blue}} = 1 - f^{\mathrm{red}}$, $f^\mathrm{red}_\mathrm{cen} + f^\mathrm{blue}_\mathrm{cen} = f_\mathrm{cen}$ of the entire sample. Note that these rescalings are necessary when converting between any two samples with different characteristics.

\begin{figure}
\centering
\includegraphics[width=\columnwidth]{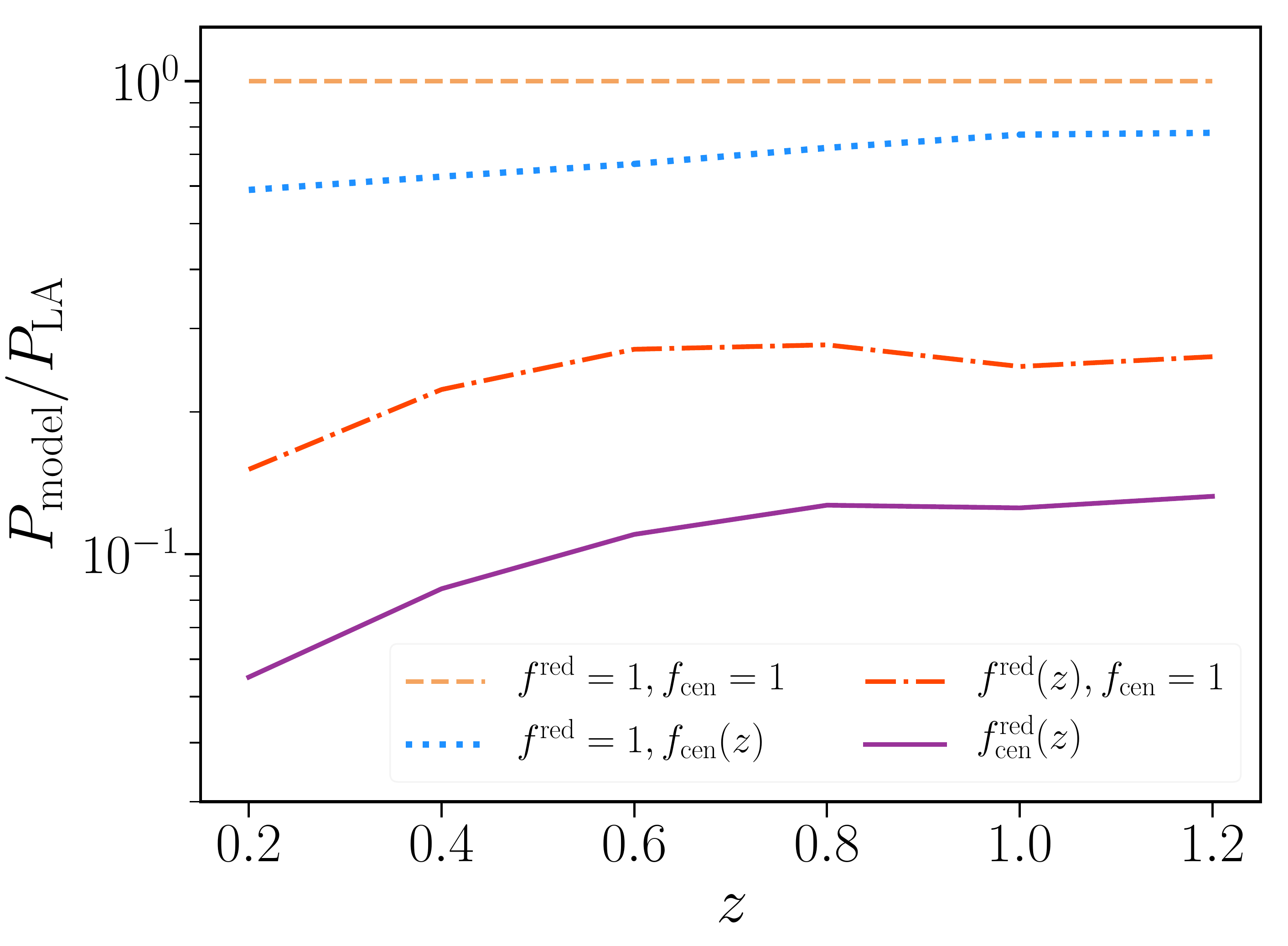}
\caption{An illustration of the redshift dependence of the IA power spectrum at large scales (2-halo regime) due to the change of the fraction of red and satellite galaxies over the $z$-bins for our simulated cosmic shear survey. We plot the ratio of a `weighted' GI power spectrum and the standard LA one. We assume a constant signal with amplitude $A=1$ (gold dashed line); incorporating the satellite fraction decreases the overall amplitude; at high redshift the fraction of satellites drops (see table \ref{tab:zbin_galinfo}), with a consequent increase of the signal (blue dotted line). At high redshift blue galaxies become important, suppressing the signal (red dot-dashed line). In this toy model, only red central galaxies are expected to contribute to the total signal (purple solid line).}
\label{fig:AI_z_cen_fractions_log}
\end{figure}

Figure \ref{fig:AI_z_cen_fractions_log} shows the competing effect of blue and satellite galaxies in suppressing the signal at high and low redshift respectively. Assuming a constant IA signal with amplitude $A=1$, it illustrates how the change of the red and the satellite fractions across the tomographic bins can affect the IA amplitude at large scales. The purple solid line shows the total amplitude (equation \ref{eq:LA_fraction_weighted_GI}): the evolution of the fraction of red central galaxies induces a sample-dependent redshift evolution of the \emph{measured} IA signal. Weighting the signal by the (red) central galaxies only significantly reduces the amplitude of the predicted IA.

\subsection{Luminosity dependence of the IA signal} \label{subsec:a_weighted_linear_alignment}

A luminosity dependence of the IA signal has been explored in the context of the large scale alignment of elliptical galaxies in a number of works \citep[][\citetalias{Johnston2019}]{Hirata2007, Joachimi2011b, Singh2015}. A common approach to model this is to follow the parametrisation in \citet{Joachimi2011b}:
\begin{equation}\label{eq:lum_dep}
    A_\mathrm{IA} \mapsto A_{\beta} \left( \frac{L}{L_0} \right)^{\beta}
\end{equation}
where $L_0$ is a pivot luminosity, assumed to correspond to $M_r=-22$. 

The value of $\beta$ is, however, being debated: while \citet{Joachimi2011b} in the MegaZ-LRG + SDSS LRG + L4 + L3 samples - hereafter simply MegaZ - and \citet{Singh2015} in LOWZ find similar values ($A_\mathrm{MegaZ} = 5.76^{+0.60}_{-0.62}$ $\beta_\mathrm{MegaZ} = 1.13^{+0.25}_{-0.27}$; $A_\mathrm{LOWZ} = 4.5^{+0.6}_{-0.6} $, $\beta_\mathrm{LOWZ} = 1.27 ^{+0.27}_{-0.27}$), \citetalias{Johnston2019}, fitting to red galaxy alignments in the GAMA + SDSS Main samples, find $A_\mathrm{G+S} = 3.17^{+0.55}_{-0.54}$ and $\beta = 0.09^{+0.32}_{-0.33}$\footnote{ Mismatched definitions of the pivot luminosity $L_{0}$, between SDSS Main and GAMA, prompted us to recompute MCMC chains for the $A_{\beta},\beta$ constraints, such that they differ slightly from those reported in \citetalias{Johnston2019} -- the updated constraints are consistent in all cases and conclusions from that work remain unchanged.}. As pointed out by \citetalias{Johnston2019}, the galaxies employed in their study contain a larger fraction of satellites compared to the MegaZ and LOWZ samples; the way this can impact the luminosity dependence is, however, non-trivial. 

\begin{figure}
\centering
\includegraphics[width=\columnwidth]{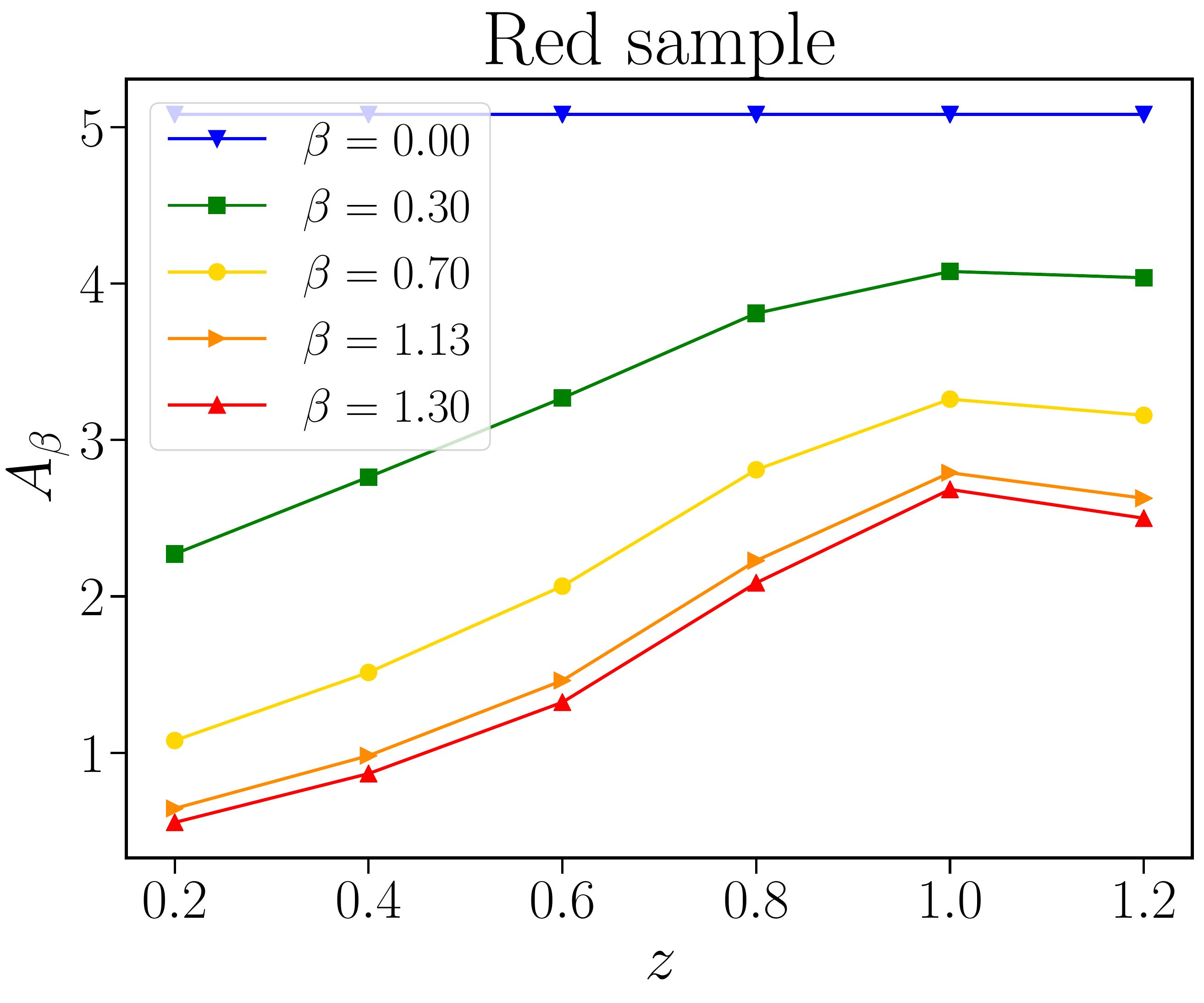}
\caption{The IA signal for different values of the slope of the power law $\beta$. The observed $z-$dependence of the signal is only caused by the different galaxy samples that populate the redshift bins. Here, we only consider the luminosity of the red central population in our simulation.}
\label{fig:A_IA_z_varying_beta}
\end{figure}

Since cosmic shear surveys span a range in luminosity much broader than what is used in those analyses, the impact of a luminosity dependence can be important in modulating the signal over the redshift bins. Different values of $\beta$ can lead to a significantly different contamination of lensing measurements \citep{Chisari2015Malquistbias}. To illustrate this, we consider the case of a population of red central galaxies with an input amplitude of $A_{\beta} \sim 5$, and vary the value of $\beta$ (Fig. \ref{fig:A_IA_z_varying_beta}). The typical luminosity of the different redshift tomographic bins causes a redshift dependence in the signal. We also note that the typical luminosity of the red central sample per $z$-bin in our mock survey is always below the pivot luminosity of $M_r=-22$ (Table~\ref{tab:zbin_galinfo}), such that the effect of the luminosity dependence is to reduce the effective IA amplitude. To evaluate the average luminosity scaling $\langle \left( L/L_0 \right)^{\beta}\rangle$ for our comic shear-like sample, we integrate over the pdf shown in Fig. \ref{fig:lum_pdf}b. 

Since the impact of an evolving signal would not be captured by a fixed IA amplitude, as often done in weak lensing analyses, it is important to understand whether a luminosity dependence exists in the data. A way around is to introduce a $z-$dependence in the alignment model, which can effectively capture the alignment variation across the $z-$tomographic bins. We do not include any \textit{intrinsic} redshift evolution in our IA model as, currently, there are only weak constraints on it \citep{Joachimi2011b,Samuroff2018DES} but we consider its effectiveness in capturing the sample-induced redshift dependence in Sect.~\ref{sec:results}.

We point out that since the luminosity dependence has only been observed in the context of red galaxy alignment, in the rest of this section we limit the discussion to the red population only.

\subsubsection{The case of GAMA galaxies}\label{subsec:the_case_of_gama_galaxies}

\begin{figure}
\centering
\includegraphics[width=\columnwidth]{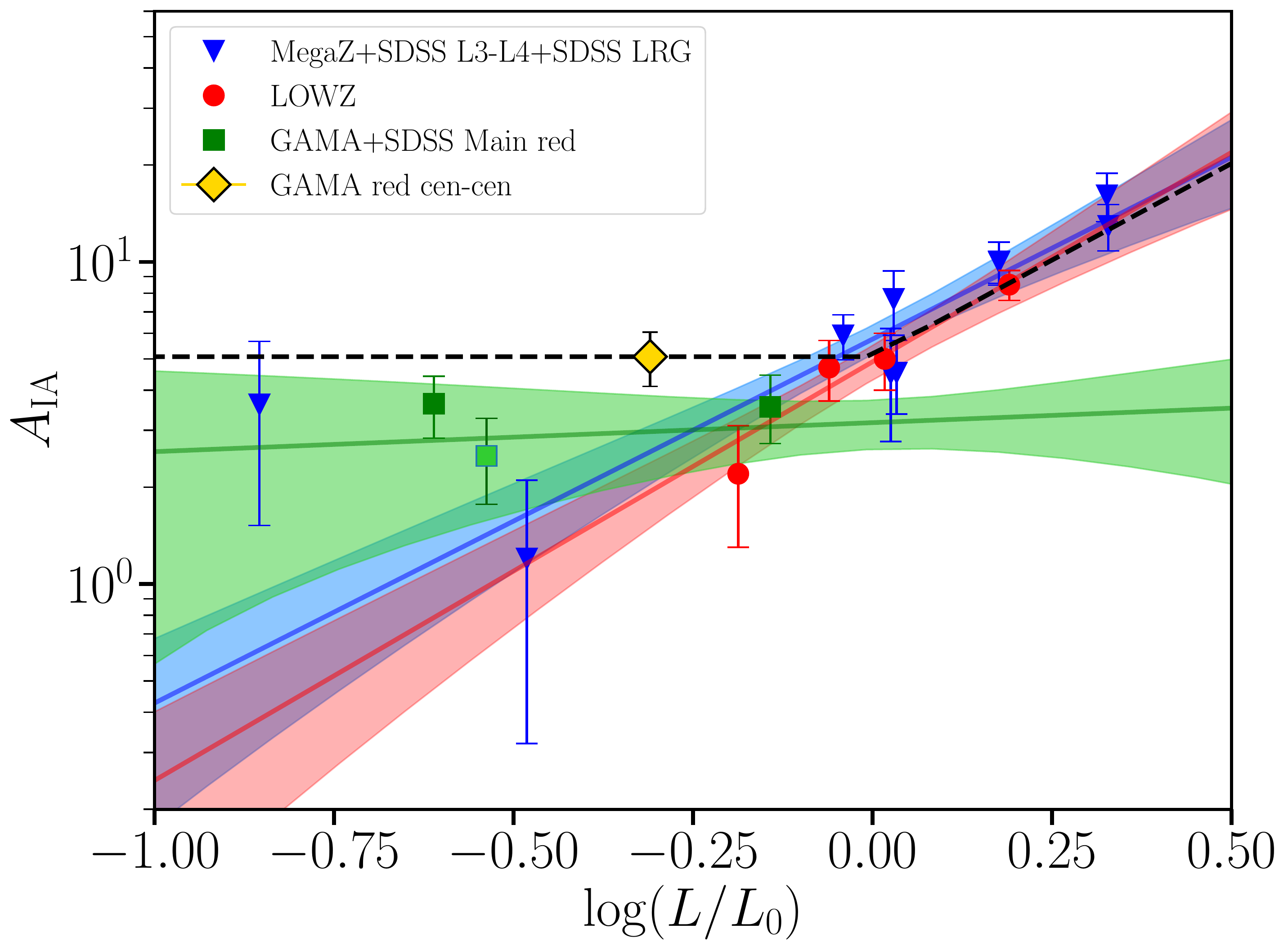}
\caption{Overview of different estimates of the IA amplitude as a function luminosity. The best-fit relation from \citet[][]{Joachimi2011b} (blue line) for the MegaZ, SDSS L3 and L4 and SDSS LRG samples (blue downward facing triangles);  \citet[][]{Singh2015} best-fit (red line) on LOWZ (red circles) and the revised best-fit to GAMA+SDSS Main from \citetalias{Johnston2019} reported in the text (green line). The three individual samples used by \citetalias{Johnston2019} are shown as green squares (GAMA) and limegreen (SDSS Main sample). The yellow diamond indicates our best fit amplitude for the GAMA red central galaxies.}
\label{fig:A_L_gama_cencen}
\end{figure}

The GAMA survey is a highly complete  spectroscopic survey (>98 per cent in the $r$-band down to $r=19.8$), which overlaps with $\sim 180$ deg$^2$ of KiDS data. The KiDS data provide high-quality galaxy images, from which \citet[][]{Georgiou2019} has measured the shapes with the DEIMOS (DEconvo-lution In MOments Space) shape algorithm \citep{Melchior2011}. This shape catalogue is employed in \citetalias{Johnston2019} and \citetalias{Georgiou2019b} for IA studies.

\citetalias{Johnston2019}'s fit of $\beta$ is obtained using three samples of red galaxies: the SDSS Main, and two samples from the GAMA, cut at $z=0.26$ in two equally populated redshift bins, Z1 and Z2. The individual fits to these samples are shown in Fig. \ref{fig:A_L_gama_cencen}. In this section, we explore whether the discrepancy on the value of $\beta$ can be due to the presence of satellites in their samples. We focus on the GAMA samples only, for which we can obtain an estimate of the fraction of satellites through the GAMA Group Catalogue\footnote{\url{http://www.gama-survey.org}}. 

GAMA ($h=0.7$) data-points from \citetalias{Johnston2019} must be shifted in the ${\rm{log}}(L/L_{0})$ axis by a factor $h^{-2}$ in order to align conventions for the pivot luminosity $L_{0}$ with SDSS ($h=1$); a re-analysis of the \citetalias{Johnston2019} luminosity dependence, with $L_0$ convention homogenised for all of their samples, does not significantly change the slope of their best-fit relation\footnote{We also note that the fiducial $\sigma_8$ in \citetalias{Johnston2019} was
misquoted as 0.8, and should in fact be 0.73. Their IA model constraints are unaffected, though their best-fit galaxy biases should be rescaled by $0.73/0.8 \sim 0.91$ to compare with a $\sigma_8=0.8$ cosmology.}. Here, we follow the $h=1$ convention, such that all the ratios reported are assuming $L_0 = 4.69\times 10^{10} L_{\odot} h^{-2}$. Interestingly, this means that the measurements from \citetalias{Johnston2019}  cover a region of the parameter space different from \citet[][]{Joachimi2011b} and \citet[][]{Singh2015}.

As mentioned in Section \ref{sec:modelling_red_central_fraction_impact}, satellite galaxies tend to randomly orient their shapes at large scales, not contributing to a alignment signal. At the same time, they preferentially lie along the major axis of their central galaxy, contributing to the satellite position - central shape correlation. In a halo model fashion, we can think of any possible contribution to sum up linearly (i.e. central position - central shape, central position - satellite shape, satellite position - central shape, satellite position - satellite shape), weighted by the fraction of galaxies that contribute to each term, together yielding the final signal that we measure. 

The individual fits to the alignment signals of the Z1 and Z2 samples in \citetalias{Johnston2019} show roughly a similar amplitude ($A_\mathrm{Z1} = 3.63^{+0.79}_{-0.79}$, $A_\mathrm{Z2} = 3.55^{+0.90}_{-0.82}$) corresponding to galaxies of different luminosity ($\langle L/L_0 \rangle_{Z1} = 0.25$, $\langle L/L_0 \rangle_{Z2} = 0.72$), compatible with their finding of no luminosity dependence. However, at low redshift the fraction of satellite galaxies in their red population is roughly $0.36$, which decreases to $f_\mathrm{sat} \sim 0.27$ in the second redshift bin.

We have seen in Sect.~\ref{sec:from_observations_to_models_satellites} that the net effect of satellites is to lower the measured amplitude. To get a sense of how this might affect our data points, we up-weight the signal by the fraction $f_\mathrm{sat}$ in each given bin: this increases the signal, which maintains the same flat relation, without significant tilts. We note that this up-weighting procedure is not quite correct and therefore should not be considered as the underlying \textit{true} shape signal, because the gI correlation contains two terms that suppress the amplitude (i.e. those for which the satellites act as shape tracers) and two where they contribute positively to the final amplitude (the central-central correlation and the satellite position - central shape correlation). Our re-weighting does not consider the positive contribution of the satellite position - central shape correlation and thus overestimates the suppression induced by the satellites. Nevertheless, it gives us a sense of the overall shift and can be considered as an upper-limit to the expected central-only alignment amplitude.

To further explore the role played by the satellites, we measure the IA amplitude of the red central sample only in GAMA ($cc$ correlation). The mean luminosity of this sample is $\langle L/L_0 \rangle = 0.46 $, for which we find a best fit amplitude $A_{\mathrm{GAMA},cc} = 5.08^{+0.97}_{-0.95}$, with a reduced $\chi^2 = 2.0$ ($N_\mathrm{dof}=4$). 

This measurement does not agree with the curve predicted by MegaZ and LOWZ, which would correspond to $2.40^{+0.59}_{-0.47}$ at that given luminosity (assuming MegaZ best-fit parameters), as illustrated in Fig. \ref{fig:A_L_gama_cencen}. Our new measurement is displayed as a yellow diamond, while the predicted best fit luminosity dependent IA amplitude measured by MegaZ and LOWZ are shown as blue and red curves, respectively. Note that the MegaZ best fit curve also includes a $z-$dependent power law that was poorly constrained in that work. We do not include it here, as recent studies have not found evidence for an intrinsic $z$-dependence of alignment strength, so the curve reported in Fig. \ref{fig:A_L_gama_cencen} is only the luminosity dependent part of their fit.

\subsubsection{A central-only luminosity dependent signal} \label{subsec:a_central_only_luminosity_dependence}

The complexity of the arising picture does not allow for a direct interpretation of the role of satellites in the context of the luminosity dependence, but we can at least identify two main scenarios.

\begin{enumerate}
    \item Central galaxies follow a single power law as observed in \citet{Joachimi2011b} and \citet{Singh2015} on MegaZ and LOWZ galaxies. The lack of such luminosity dependence detection in \citetalias{Johnston2019} can be accounted by the non-negligible presence of satellites in their sample. The fact the measurement of the central-central galaxy alignment from GAMA does not coincide
    with the MegaZ/LOWZ predictions can point towards a shallower relation than what was measured by those samples. This can be a consequence of satellites also contaminating the MegaZ and LOWZ samples.
    \item Bright central galaxies follow the luminosity dependence in \citet{Joachimi2011b} and \citet{Singh2015}, while faint galaxies are characterised by a different slope, in a double power law scenario. Given the current measurements in this part of the parameter space, the most extreme case is a flat luminosity dependence for $L<L_0$ ($\beta_{L<L_0}=0$).
\end{enumerate}

In all of these cases, we are restricting the IA luminosity dependence at large scales to central galaxies, a choice that finds a natural theoretical frame in the context of the linear alignment mechanism, where the intrinsic shear power spectrum can be expressed as a power of the mass of the hosting halo \citep{Piras2018}. This can in turn be related to the luminosity of its central galaxy. In the rest of the paper, we assume that in the 2-halo regime, the luminosity dependence is only caused by the central galaxy population and that the bright-end of such relation is well described by the best-fit values from MegaZ and LOWZ analyses. At the faint-end, we allow for both scenarios described above: our case (i) corresponds to the luminosity dependence from MegaZ and LOWZ, such that the luminosity dependence is described by a single power law, 
\begin{equation}\label{eq:lum_dep_modified_GI}
    P^\mathrm{red}_{\mathrm{\delta I}}(k,z,L) = f^\mathrm{red}_\mathrm{cen} P_{\delta \mathrm{I}}(k,z)  \Biggl\langle \left( \frac{L^\mathrm{red}_\mathrm{cen}}{L_0} \right)^{\beta} \Biggr\rangle
\end{equation}
and
\begin{equation}\label{eq:lum_dep_modified_II}
    P^\mathrm{red}_{\mathrm{II}}(k,z,L) = (f^\mathrm{red}_\mathrm{cen})^2 P_{\mathrm{II}}(k,z)  \Biggl\langle \left( \frac{L^\mathrm{red}_\mathrm{cen}}{L_0} \right)^{\beta} \Biggl\rangle^2 \ ,
\end{equation}
while in our case (ii) we consider a broken power law that passes through the best fitting amplitude of the central-central correlation, flat for $L<L_0$ and then assuming the form of a power law with index $\beta=1.2$ for $L>L_0$. This is shown with a black dashed line in Fig, \ref{fig:A_L_gama_cencen}. Any intermediate slope of the luminosity dependence of faint galaxies would be in between these two cases, and thus we can have an estimate of the range of impact of this term in the lensing contamination from IA. 

Although the alignment of central galaxies with luminosities $L \ge L_0$ seems to be captured better by the LOWZ/MegaZ best fit curve, we caution that LOWZ and MegaZ are not pure central galaxy samples (for example, LOWZ has roughly a fraction 11$\%$ of satellites, see \citet{Singh2015}, Sect.~3.3), which also contaminates the results, particularly at the faint end of the curve. At low luminosities, the contribution of satellite galaxies to the final signal is more important, as satellites are predominantly faint. This implies that as we move from left to right in Fig. \ref{fig:A_L_gama_cencen}, we observe a simultaneous increase of the IA signal due to the increase of the galaxy luminosity and a depletion of the satellite suppressing contribution. If we assume that MegaZ and LOWZ measurements also suffer from the presence of satellites in their low luminosity bins, the net effect would be a lower value of $\beta$, and thus a less steep relation.

This scenario remains possible as we lack a proper normalisation for this term. As bright galaxies are in general not abundant and at high luminosities the sample is not significantly contaminated by satellites, this scenario is only relevant in the intermediate luminosities around $L_0$. Below $L_0$ this falls between the two cases we are considering. Future IA studies that aim to constrain the IA signal using observations and simulations should focus on the impact of satellites on the value of $\beta$, as galaxies at $L/L_0 \sim 1$ are already expected to have $\ge 10\%$ of satellites (based on $f_\mathrm{sat}(L)$ in our GAMA sample).

\subsection{Colour dependence}

A key aspect of our approach is that we weight the alignment signal by the fraction of galaxies that contributes to that specific amplitude. We have seen that the IA alignment is strongly morphology dependent: this implies that the weighting by the red fraction plays a significant role in the prediction of the final signal. The different measurements compared in the previous section have been measured on samples selected with different red cuts. However, \citet[][]{Singh2015} has explored the dependence of the IA signal on colour, finding no evidence for a colour-dependence in the data. Since \citetalias{Johnston2019} provide IA amplitudes for both the red and blue samples, and we have shown that their measurements agree with the results in MegaZ and LOWZ once restricting the analysis to the central sample only, we adopt \citetalias{Johnston2019} amplitude for the blue population and consistently apply a red cut similar to the one in their work (see Sect.~\ref{subsec:galaxy_mocks}). 

\section{The impact of satellites at small scales} \label{sec:the_impact_of_satellites_at_small_scales}

\begin{figure}
\centering
\includegraphics[width=\columnwidth]{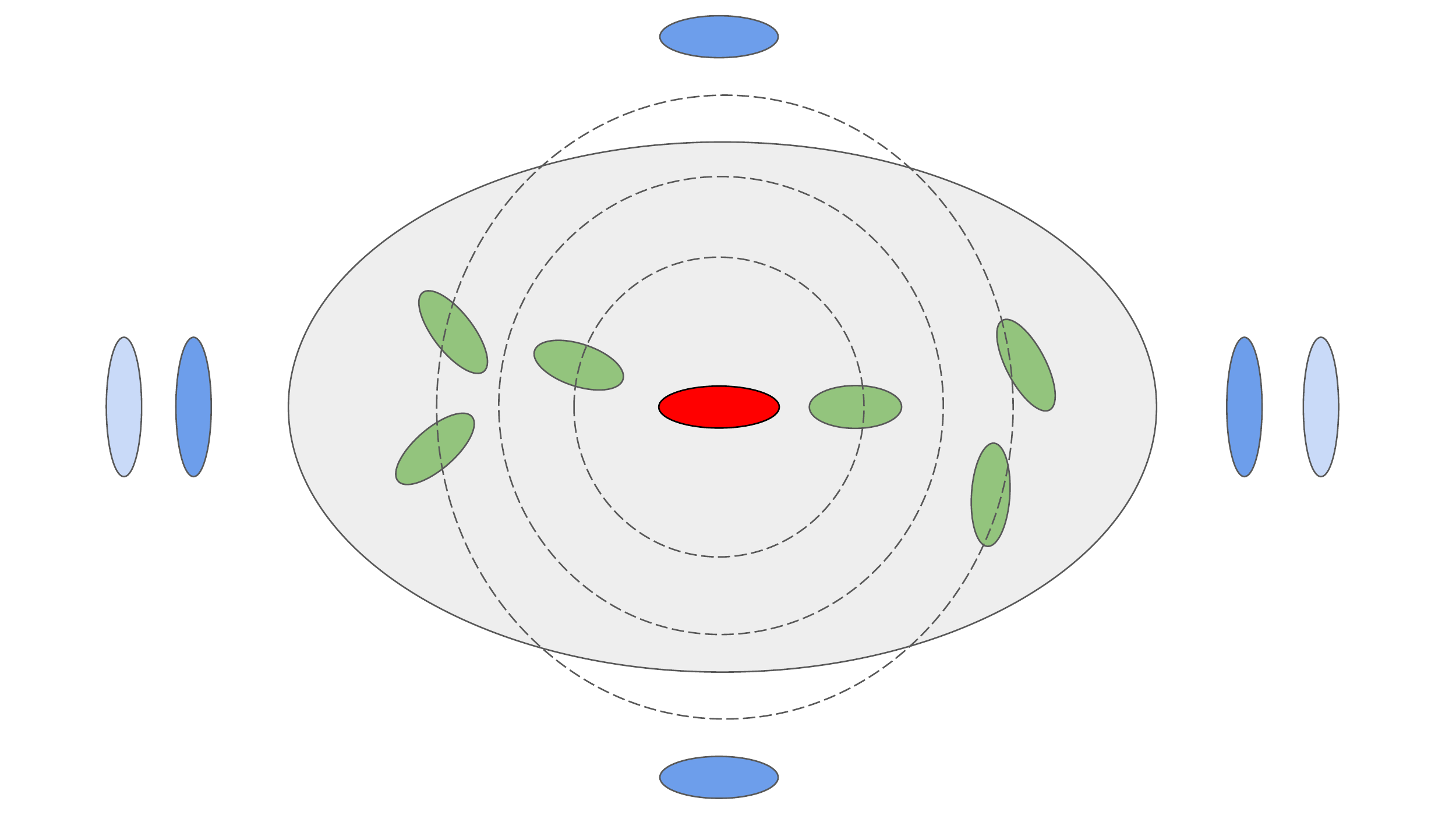}
\caption{A cartoon showing the current picture of satellite alignment provided by observations. Satellite galaxies (green ellipses) tend to preferentially segregate along the direction of the central galaxy (red ellipse) major axis; the closest satellites to the centre of the halo show a preferential alignment in the direction of the central galaxy major axis. The source galaxies (blue ellipses) are tangentially aligned with respect to the halo shape, resulting in an opposite alignment with respect to the aligned satellites and the central galaxy. We can expect source galaxies to be lensed more along the halo major axis, due to the excess of matter in that direction. The dotted circles illustrate the way a spherical halo model can describe this alignment signal.}
\label{fig:IA_sc_contamination_cartoon}
\end{figure}

To model the impact of satellite alignment at small scales, we revisit the halo model formalism by \citetalias{SchneiderBridle2010} to take into account new observational results. The spherical halo approximation should capture most of the small-scale GI signal. As discussed in Sect.~\ref{sec:modelling_red_central_fraction_impact}, the anisotropic distribution of satellites boosts the signal of the gI correlation, so a spherical model would underestimate $w_{g+}$. How it propagates exactly in the context of the GI contamination is not trivial. The satellite segregation along the central galaxy major axis is expected to source a large 1-halo satellite position - satellite shear correlation, confirmed in \citetalias{Johnston2019}, but also of an opposite satellite-satellite II term, for which we do not have any observational measurement. Moreover, if only the innermost satellites are aligned in the direction of the central galaxy, as observed in \citetalias{Georgiou2019b}, the impact on GI should be significantly reduced compared to gI. For an illustration of these terms, see the cartoon in Fig. \ref{fig:IA_sc_contamination_cartoon}. Although further study is needed, we use the spherical halo model formalism because we expect it to capture the leading contribution, providing a fair sense of the amplitude of the satellite alignment.

\subsection{The halo model formalism for satellite alignment}

Following \citetalias{SchneiderBridle2010} we adopt an effective radial satellite alignment and describe galaxy orientations inside the halo through the stick approximation. In this approximation, the two minor axes of the elliptical galaxy have equal lengths on average and the length and the orientation of the stick correspond to those of the galaxy major axis. 

Defining a Cartesian reference system centred on the halo and with the $z-$axis along the line of sight, the position of a satellite galaxy inside the halo is identified by the vector $\mathbf{r}=(r,\theta, \phi)$. The orientation of the satellite major axis can then be expressed through the unit vector 
\begin{equation}\label{eq:e_hat}
    \hat{e} = (\sin \theta_e \cos \phi_e, \sin \theta_e \sin \phi_e, \cos \phi_e),
\end{equation}
where $\theta_e$ and $\phi_e$ are, respectively, the polar and azimuthal angles that the satellite major axis forms with the radial vector $\mathbf{r}$.

In principle, we expect satellite galaxies to follow a distribution of angles between their major axes $\hat{e}$ and the radial vector $\mathbf{r}$. However, \citetalias{SchneiderBridle2010} showed that the main effect of including this term is simply to reduce the amplitude of the correlation functions with respect to the case of perfect radial alignment, independently of the halo mass. Thus, we can simply consider the case of perfect radial alignment, absorbing any misalignment into the amplitude of the intrinsic alignment signal.
In this case, $\theta_e = \theta$ and $\phi_e = \phi$. In a sense, the perfect radial alignment configuration can be considered as an effective description: we can only measure the tendency of galaxies to point in a certain direction, so the length of the sticks -- which determines the amplitude of the signal -- quantifies the combination of the amplitude of the misalignment angle and the intrinsic ellipticity of the galaxy. This provides a direct map between the formalism of the stick model and the measured alignment $|\epsilon| \langle \cos (2 \phi) \rangle$, where $|\epsilon|$ is the modulus of the ellipticity and $\phi$ is the misalignment angle.

Calling $\bar{\gamma}$ the length of the stick, and assuming the alignment to be a function of the distance to the halo centre and the mass of the halo, it follows that (\citetalias{SchneiderBridle2010})
\begin{equation}\label{eq:gamma_I}
    \gamma^I (\mathbf{r},M,c) = \bar{\gamma}(r,M,c) \sin \theta e^{i 2 \phi}.
\end{equation}
Here, $\bar{\gamma}(r,M)\sin \theta$ is the observed length of the stick, corresponding to the projection of the major axis along the line of sight. In principle, this quantity can also depend on the halo concentration, but we assume a deterministic relation between mass and concentration (see also Sect.~\ref{sec:halo_model_setup}) and so we omit such dependence in the following.

Since we only measure the IA signal at galaxy locations, it is necessary to introduce a density weighting in the model \citep{Hirata2004}, $\tilde{\gamma}^I = \gamma^I(1+\delta_g)$. Following \citetalias{SchneiderBridle2010}, we weight the 3D projected ellipticity by the number of galaxies inside the halo, $N_g$ and the normalised matter density profile $u(\mathbf{r}|M) = \rho(r|M)/M$:
\begin{align} \label{eq:sat_ellipticities}
    \tilde{\gamma}_\mathrm{1-halo}^I (\mathbf{r}, M) =
    \bar{\gamma} (r, M) \sin \theta \e^{2i \phi} N_g u(\mathbf{r}|M),
\end{align}
where we identify the density-weighted shear with a tilde \citep[][\citetalias{SchneiderBridle2010}]{Hirata2004}.

Having defined the density-weighted ellipticity $\tilde{\gamma}^I$ for a given halo, we can construct a continuous intrinsic ellipticity field by summing up the contributions from each individual halo $i$, in the usual halo model fashion:
\begin{align*}
    \tilde{\gamma}_s^I (\mathbf{r}) &= \frac{1}{\bar{n}_g} \sum_i \gamma^I(\mathbf{r}- \mathbf{r_i}, M_i) N_{g,i} u(\mathbf{r-r_i},M_i) \\
    &= \sum_i \int \dd M \ \int \dd^3 r' \ \delta_D (M-M_i) \delta_D^{(3)} (\mathbf{r-r_i}) \frac{N_{g,i}}{\bar{n}_g} \\ & \ \ \times \gamma^I (\mathbf{r-r'},M) u(\mathbf{r-r'},M) \ ,
\end{align*}
where $\bar{n}_g$ is the galaxy number density per unit of volume, which is a function of redshift. The subscript $s$ indicates that this density weighted shear only refers to satellites.

We calculate the correlation functions of interest for IA by correlating $\tilde{\gamma}_s^I (\mathbf{r})$ with itself and with the matter density contrast $\delta_m$. In Fourier space, the $E$ and $B$ modes of the IA are defined as 
\begin{equation}\label{eq:gamma_E}
    \tilde{\gamma}^I_E(\mathbf{k}) = \cos (2 \phi_k) \tilde{\gamma}_1^I(\mathbf{k}) + \sin (2 \phi_k) \tilde{\gamma}_2^I(\mathbf{k})  
\end{equation}
\begin{equation}\label{eq:gamma_B}
    \tilde{\gamma}^I_B(\mathbf{k}) = \sin (2 \phi_k) \tilde{\gamma}_1^I(\mathbf{k}) - \cos (2 \phi_k) \tilde{\gamma}_2^I(\mathbf{k})  \ , 
\end{equation}
where 
\begin{equation}\label{eq:gamma_fourier_transform}
    \tilde{\gamma}^I(\mathbf{k},M) \equiv \int \dd ^3 \mathbf{r} \ \tilde{\gamma}_j^I(\mathbf{r}, M) \e^{i \mathbf{k} \cdot \mathbf r}.
\end{equation}
is the Fourier transform of the complex density-weighted shear, with $j=1,2$ being the two components. Thus,
\begin{equation}\label{eq:from_gammaE_to_P_EE}
\langle \tilde{\gamma}_E^{I*}(\mathbf{k},z) \tilde{\gamma}_E^I(\mathbf{k'},z) \rangle = (2 \pi)^3 \delta_D^{(3)}(\mathbf{k}- \mathbf{k'}) P_{\tilde{\gamma}^I}^{EE}(\mathbf{k},z)
\end{equation}
and 
\begin{equation}\label{eq:from_delta_gammaE_to_P_deltaE}
\langle \delta^{*}(\mathbf{k},z) \tilde{\gamma}_E^I(\mathbf{k'},z) \rangle = (2 \pi)^3 \delta_D^{(3)}(\mathbf{k}- \mathbf{k'}) P_{\delta,\tilde{\gamma}^I}(\mathbf{k},z)
\end{equation}

Without loss of generality, we can rotate our reference system such that $\gamma^I_1 = \gamma^I_{+}$, where $\gamma^I_{+}$ is the tangential component of the shear. This corresponds to fixing $\phi_k=0$ in equations \ref{eq:gamma_E} and \ref{eq:gamma_B}, transforming $EE$ into II.

For computational reasons, it is convenient to separate the radially dependent part of the density-weighted shear, which is affected by the Fourier transform, from the terms that are only mass dependent. We then define:
\begin{equation}
    \hat{\gamma}_s^I(\mathbf{k},M) \equiv \mathcal{F}\left( \gamma^I(\mathbf{r},M) \ u(\mathbf{r},M) \right).
\end{equation}

We now have the ingredients to compute all of the possible IA power spectra. In a spherical halo model, the only terms that survive are: the II satellite-satellite power spectrum and the satellite-matter term for the $\delta$I power spectrum. These can be written as
\begin{equation}\label{eq:PGI_1halo_term}
    P^s_{\delta \mathrm{I, 1h}}(\mathbf{k},z) = \int \dd M \ n(M) \frac{M}{\bar{\rho}_m} f_s(z) \frac{\langle{N_s|M}\rangle }{\bar{n}_s(z)} |\hat{\gamma}^I_s(\textbf{k}|M)|  u(k,M)
\end{equation}
and
\begin{equation}\label{eq:PII_1halo_term}
    P_\mathrm{II, 1h}^{ss}(\mathbf{k},z) = \int \dd M \ n(M) f_s^2(z) \frac{\langle{N_s(N_s-1)|M}\rangle }{\bar{n}^2_s(z)} |\hat{\gamma}^I_s(\textbf{k}|M)|^2
\end{equation}
where $n(M)$ is the halo mass function, $f_s(z)$ is the fraction of satellite galaxies as a function of redshift and $\langle N_s|M \rangle$ is the halo occupation distribution of satellite galaxies.

The power spectra in equations \ref{eq:from_gammaE_to_P_EE}-\ref{eq:from_delta_gammaE_to_P_deltaE} are functions of $(k, \theta_k)$. However, $\theta_k$ only modulates the strength of the amplitude of the signal. In the rest of the paper, we decide to fix $\theta_k=\pi/2$ (Limber approximation; for a more detailed discussion of the angular dependence of the power spectra, see Appendix \ref{A:angular_terms_in_IA}). 

\subsection{Radial dependent satellite alignment}
\label{subsec:radial_dependent_satellite_alignment}

We use the mean radial alignment signal $\langle \epsilon_+ \rangle$ measured in \citetalias{Georgiou2019b} to model the satellite alignment in Eq. \ref{eq:sat_ellipticities}, $\bar{\gamma}(\mathbf{r},M)$. They measured a satellite alignment in bins of projected distance of the satellite from the group's brightest galaxy, $r_\mathrm{sat}/r_{200}$ and found a radially dependent signal. It is well-fitted by a power law of the form $\langle \epsilon_+ \rangle = A(r_\mathrm{sat}/r_{200})^b$. The slope is chosen to be fixed at $b=-2$ and the amplitude is fit for the different galaxy samples. \citetalias{Georgiou2019b} do not detect any mass dependence, so we do not include it in our parametrisation.

The estimator $\langle \epsilon_+ \rangle$ is related to the shear via
\begin{equation}\label{eq:e2gamma}
    \gamma_+ = \frac{\langle \epsilon_+ \rangle}{\mathcal{R}}  \ ,
\end{equation}
where $\mathcal{R}$ is the shear responsivity and quantifies the response of the ellipticity to a small shear\footnote{Our definition of ellipticity is $|\epsilon| = (1-q)/(1+q)$, where $q$ is the semi-minor to semi-major axis ratio and thus we do not need to double the responsivity in eq. \ref{eq:e2gamma}, as in the case of shapes measured via polarisation.}. Rounded object are easier to shear than highly elliptical objects. Here, we assume a typical value of $\mathcal{R} \approx 1 - \sigma_{\epsilon}^2 = 0.91$ to convert $\epsilon_+$ to $\gamma^I$.

In order to prevent unphysiscal behaviour at very small scales, we adopt a piecewise function of the form
\begin{equation}\label{eq:radial_dependence_piecewise}
    \bar{\gamma}(r)=\begin{cases} a_{1h} \left( \frac{0.06}{r_\mathrm{vir}} \right)^b, & \mbox{if } r<0.06 \mathrm{\ Mpc}/h  \\ a_{1h} \left( \frac{r}{r_\mathrm{vir}} \right)^b, & \mbox{if } r>0.06 \mathrm{\ Mpc}/h  \ ,
    \end{cases}
\end{equation}
where $a_{1h}$ is the amplitude of the power law. We further impose that $\bar{\gamma}(r)$ never exceeds 0.3, which corresponds to a perfect alignment. We choose to set $r= 0.06$ Mpc/$h$ based on the minimum angular separation for a shape measurement, which we assume to be $\theta_\mathrm{max} = 4$ arcsec for a ground-based telescope. At high redshifts, the largest separation that can be resolved is around 60 kpc$/h$. At low redshifts, the spatial resolution is much smaller, but the light coming from the central galaxy, in particular for the most massive ones, can contaminate the measurements up to this scale \citep{Sifon2018}. A cut at $0.06$ Mpc/$h$ slightly suppresses the signal of the low mass galaxies, for which the satellite alignment is expected to be small. A space-based telescope such as \textit{Euclid} will be able to resolve objects down to a smaller separation, but the physical extent and the contamination from the central galaxy still impose a truncation at small scales. Reducing the transition scale increases the amplitude of the signal, as satellites get more and more aligned as we approach the centre of the group/cluster. We experimented with different values of the truncation parameter and we find the impact to be subdominant with respect to the other source of uncertainties considered in this paper. However, future lensing studies that aim to include very small separations have to cope with an increasing IA contamination, in a regime where we do not have observations to properly calibrate its impact. 

The measurements by \citetalias{Georgiou2019b} are performed in projection, which means that we can easily relate the projected shear $\gamma^I$ to $\langle \epsilon_+ \rangle$. However, the radial position within the halo in our framework requires the 3d position, while \citetalias{Georgiou2019b} measure the signal in projected distance. For $r>0.06$ Mpc$/h$, this introduces a $\sin^b \theta$ term in our expression of $\gamma^I$ (see also Appendix~\ref{A:angular_terms_in_IA}),
\begin{equation}\label{eq:gamma_I_sin_theta}
    \gamma^I(r, \theta) = \bar{\gamma}(r) \sin \theta = a_\mathrm{1h} \left( \frac{r \sin \theta}{r_\mathrm{vir}} \right)^b \ .
\end{equation}

As for the large scale signal, we distinguish between the alignment of red and blue satellites, for which \citetalias{Georgiou2019b} find a different amplitudes.

\subsection{Luminosity dependence of the satellite galaxy alignment}\label{subsec:luminosity_dependence_of_the_satellite_galaxy_alignment}

\begin{figure}
\centering
\includegraphics[width=\columnwidth]{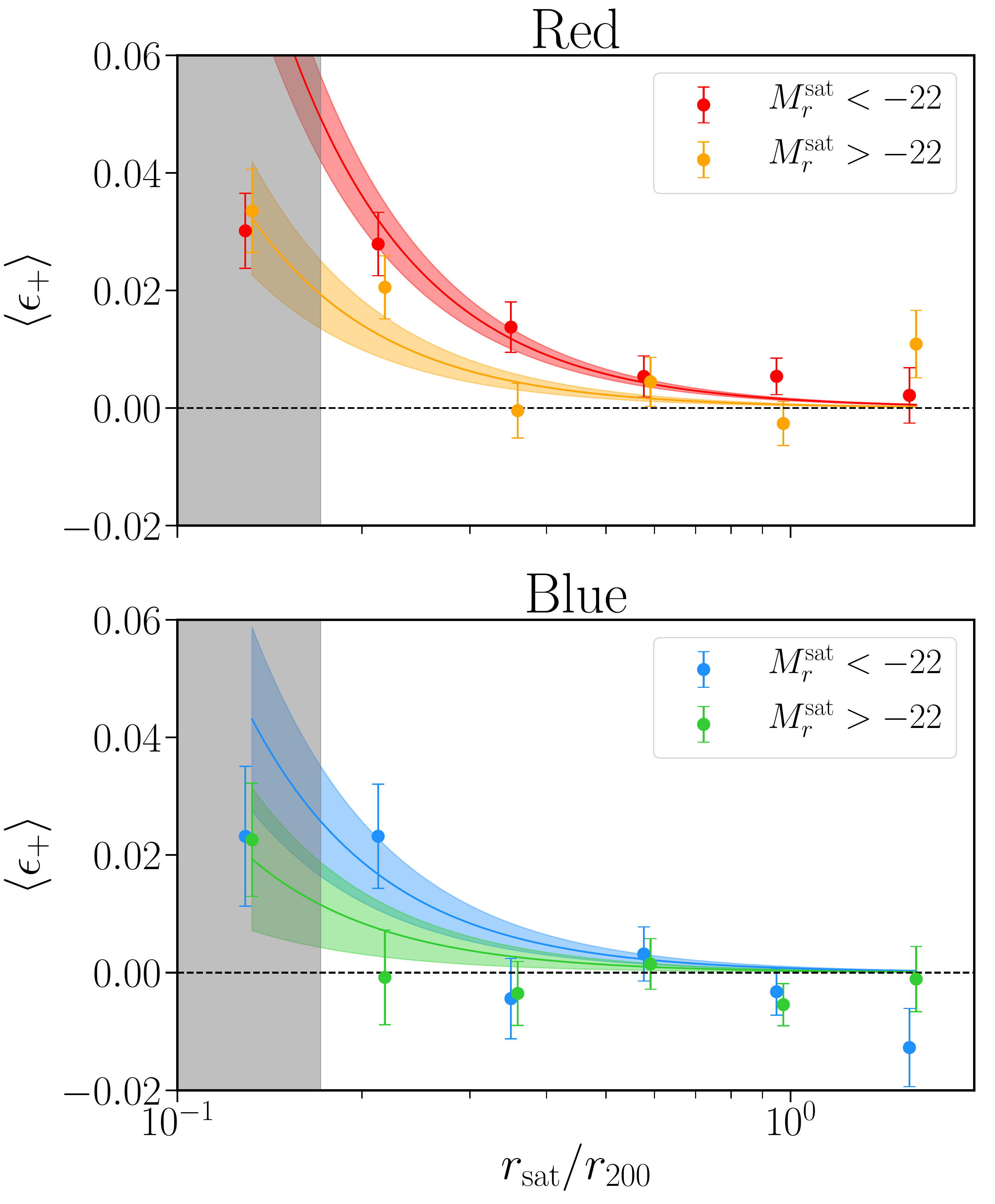}
\caption{Mean tangential ellipticity components versus satellite projected distance from the brightest galaxy in the group, for the galaxy sample in \citetalias{Georgiou2019b}. We separately consider the red (top panel) and blue (bottom panel) sample. We jointly fit the luminosity and radial dependence for each of the two samples, as in equation \ref{eq:satellite_radial_luminosity_dependence}. The data points in the grey region are excluded from the fit. Our best fit is on top of the data points, with the $1-\sigma$ uncertainty on the fit.}
\label{fig:sat_eplus_red_blue_lum_dep}
\end{figure}

In the spirit of including any observational insights into our model, we focus here on the luminosity dependence of satellites. This was detected in \citet{Huang2018} in the SDSS redMaPPer galaxies but not confirmed by \citetalias{Georgiou2019b} in galaxy groups. The same dependence was explored in clusters by \citep{Sifon2015}, who did not find any evident trend with the given S/N. 

Addressing whether a luminosity dependence of satellites exists is particularly important for the IA contamination of the lensing signal in the low redshift bins, where the satellite population is more abundant and spans a large range in luminosities. In particular, the lowest redshift bins of a typical lensing survey do not reflect the satellite population employed in \citetalias{Georgiou2019b}, containing a larger fraction of faint blue satellites, while their sample peaks at $M_r~-22$ and has an equivalent fraction of red and blue satellites. If faint satellites are characterised by a different alignment behaviour with respect to the bright ones, extrapolating their findings might largely overestimate the IA impact on such bins.

We focus on the SDSS-redMaPPer and GAMA+KiDS analyses here. \citet{Huang2018} observed that when using the re-Gaussianization shape algorithm, the satellites with $^{0.1}M_r<-21$, located closer to central galaxies, show a more prominent signal. Since the redMaPPer algorithm selects luminous red galaxies, one of the major differences between the galaxy samples used in the two studies is the colour of the satellites. \citetalias{Georgiou2019b} investigate the luminosity dependence only for the full sample, while \citet{Huang2018} focus on the red population only. 

We re-analyse the galaxies in \citetalias{Georgiou2019b}, looking for a luminosity scaling of the signal for the two separate cases of red and blue galaxies. We select red galaxies imposing the same cut as \citetalias{Georgiou2019b}. We split the samples into two bins, cutting at $M_r=-22$ to ensure that the two bins have a comparable number of galaxies. We detect a luminosity dependence for both the red and blue sample. As before, we fit a power law with fixed index $b=-2$. Our results are summarised in Table \ref{tab:sat_lum_dep_fit}. Following \citetalias{Georgiou2019b}, we do not include the first radial bin in our fits, since the light from the brightest galaxy of the group biases the shapes.

Although the signal-to-noise ratio does not allow for a definitive constraint on the luminosity dependence of the satellite alignment, we can draw the following conclusions: the faint blue satellites do not show any alignment signal, while the bright sample shows an alignment signal only for the innermost radial bin. The red satellites show a more prominent signal for both the faint and the bright samples. While the bright sample of the blue and the red satellites are still consistent with each other within the error bars, what drives the main difference in the red and blue satellite alignment is the behaviour of the faint bin.

\begin{table*}
	\caption{Satellite luminosity dependence best fit amplitude ($a_\mathrm{1h}$) for the red and blue sample. The samples are split in two luminosity bins, L1 (bright) and L2 (faint), with a cut at $M_r=-22$. In all of the fits, we assume a radial dependence with the form of a power law with slope -2, as in \citetalias{Georgiou2019b}. In the joint fit, the luminosity dependence is modelled with a power law with slope $\zeta$, as in eq. \ref{eq:satellite_radial_luminosity_dependence}.}
	\label{tab:sat_lum_dep_fit}
	\begin{tabular}{lccccr} 
		\hline
	Sample & & N$_\mathrm{gal}$ & $a_{1h}$ & $\zeta$ & $\chi^2/\mathrm{dof}$ \\
			\hline
			\hline
    \multicolumn{6}{l}{Individual sample fits:} \\
		\hline
Red & L1 & 7505 & $0.0014 \pm 0.0002$ & - & 0.59 \\
 & L2 & 6618 & $0.0008 \pm 0.0002$ & - & 1.76 \\
Blue & L1 & 5989 & $0.0008 \pm 0.0004$ & - & 1.97 \\
 & L2 & 8778 & $-0.0002 \pm 0.0003$ & - & 0.67 \\
	    \hline
    \multicolumn{6}{l}{Joint fit:} \\
        \hline
Red & all & & $0.0009 \pm 0.0001$ & $0.7 \pm 0.2$ & 1.02 \\
Blue & all & & $0.0006 \pm 0.0002$ & $0.5 \pm 0.4$ & 1.50 \\ 
    \hline
	\end{tabular}
\end{table*}

To model the luminosity dependence, we decide to follow the parametrisation adopted for the red central galaxies, a power law in $L/L_0$, where $L$ is now the luminosity of the satellite sample under consideration and $L_0$ is the pivot luminosity, corresponding to a magnitude of $M_r=-22$.  We perform a joint fit of the radial and luminosity dependence for the red and blue sample separately, assuming the functional form:
\begin{equation}\label{eq:satellite_radial_luminosity_dependence}
    \langle \eps_+ \rangle (r_\mathrm{sat}, L) = a_{1h} \left( \frac{L}{L_0} \right)^{\zeta} \left( \frac{r_\mathrm{sat}}{r_\mathrm{vir}} \right)^b \ .
\end{equation}

As for the rest of the analysis, we do not fit for $b$, which is chosen to be $b=-2$. Table \ref{tab:sat_lum_dep_fit} reports our best fit values of $a_{1h}$ and $\zeta$ for the two samples, and Fig. \ref{fig:sat_eplus_red_blue_lum_dep} shows our best fit curves on top of the data points.

\section{Halo model setup} \label{sec:halo_model_setup}

To inform our model about the properties of the galaxy sample for which we predict the IA signal, we extract the HODs of central and satellites from our Stage III survey mock. We checked that this procedure gives us number densities of galaxies that match those measured in the simulations in redshift bins.

We define dark matter haloes as spheres with an average density of $200\bar{\rho}_m$. The mass of the haloes provided by MICE is based on the Friends-of-Friends (FoF) algorithm. The two definitions slightly differ from each other, in particular at high redshifts. We employ MICE masses only when computing the HODs, which enter in the small scales of the model. Those scales are important at low redshift only, so this mass-definition discrepancy is expected to not have a major impact for our analysis. This is further confirmed by the fact that we can recover compatible measured galaxy number densities within our halo model setup. In the following, we always use the $M_{200}$ definition.

We assume that dark matter haloes follow the Navarro-Frenk-White distribution \citep{NFW1996}, with a concentration-mass relation from \citet{Duffy2008} and that satellite galaxies are spatially unbiased with respect to the dark matter particles\footnote{We do not provide galaxy positions within the halo as implemented in MICE, but only the mean halo occupation, $N_g$ given the mass of the halo.}. For the halo mass function and for the halo bias function we adopt the functional forms from \citet{Tinker2010}. For the implementation of the former we make use of the public available python package HMF\footnote{\url{https://github.com/steven-murray/hmf}} \citep{Murray2014HMF}.

The total IA power spectra are given by the sum of the introduced contributions introduced in Sect.~\ref{sec:modelling_red_central_fraction_impact} eq.~\ref{eq:lum_dep_modified_GI}-\ref{eq:lum_dep_modified_II}, describing the behaviour at large scales (2h regime), and at small scales, presented in Sect.~\ref{sec:the_impact_of_satellites_at_small_scales}, eq.~\ref{eq:PGI_1halo_term}-\ref{eq:PII_1halo_term}. When evaluating the mass integrals we consider masses in the range $[10^{10.5}, 10^{15.5}] M_{\odot}/h$ to match the observed one in our mocks. To weight the IA signal, we do not use the galaxy fractions that we directly measure from the simulations, but those computed as the integral of the HOD and the halo mass function. In this way we ensure that the large and small scales contain the same galaxy numbers and ratios. Our recovered fractions are overall more accurate for the red sample, with an error around $5\%$ in the relevant bins, while for the blue sample, we recover the true values with an error of $15\%$. As blue galaxies have a smaller alignment amplitude, this is not a major concern here. We have also tested that our main results are not affected by reasonable changes of the mass ranges.

As discussed in Sect.~\ref{subsec:a_central_only_luminosity_dependence}, to model the large-scale alignment of the red sample, we consider two cases:
\begin{enumerate}
    \item Simple power-law: $A_\mathrm{red}=5.33 \pm 0.6$ and a luminosity dependence with slope $\beta=1.2 \pm 0.4$, given by the weighted mean of LOWZ and MegaZ best fit $A_0$ and $\beta$
    \item Broken power-law: $A_\mathrm{red}=5.08^{+0.97}_{-0.95}$, $\beta_{L<L_0}=0$ and $\beta_{L\ge L_0}=1.2 \pm 0.4$
\end{enumerate}

For the blue galaxy alignment at large scales, we refer to the best fit amplitude in \citetalias{Johnston2019}, who found $A_\mathrm{blue} = 0.21 \pm 0.37$\footnote{We decided to use \citetalias{Johnston2019} best fit amplitude for the blue sample as input parameter for the IA signal of the central blue galaxies only. This is motivated by the fact that our sample is significantly fainter than the one used in \citetalias{Johnston2019} and we have seen that faint blue satellites to not show any alignment signal (Sect.~\ref{subsec:luminosity_dependence_of_the_satellite_galaxy_alignment}).}.
For the satellite alignment we consider a combined radial and luminosity dependence, as discussed in Sect.~\ref{subsec:luminosity_dependence_of_the_satellite_galaxy_alignment}. We de-project the signal as described in Sect.~\ref{subsec:radial_dependent_satellite_alignment}, so that our final $\gamma^I$ is in terms of $r$ rather than the projected separation $r_\mathrm{sat}$. The radial dependence is described by \ref{eq:gamma_I_sin_theta}, including the piece-wise term as in equation \ref{eq:radial_dependence_piecewise}. Note that in the case of the luminosity dependence, $L$ is the mean luminosity of the red/blue satellites for each redshift tomographic bin. 
The final parameters are summarised in Table \ref{tab:halo_model_setup}.  

\begin{table}
	\caption{The IA parameters adopted in our model. In the 2-halo regime, we consider two different cases for the luminosity dependence of the central galaxy population: a single power law (case i) and a double (broken) power law (case ii).}
	\label{tab:halo_model_setup}
	\centering
	\begin{tabular}{cccc}
	    \hline
	Sample & Model & Parameter & Value \\
		\hline
		\hline
    Red & 2-halo (i) & $A_\mathrm{\beta}$ & $5.33 \pm 0.60$ \\
     & & $\beta$ & $1.2 \pm 0.27$ \\
     & 2-halo (ii) & $A_\mathrm{\beta}$ & $5.08 \pm 0.97$ \\
     & & $\beta_{L\ge L_0}$ & $1.2 \pm 0.27$ \\
     & & $\beta_{L< L_0}$ & $0$ \\
     & 1-halo & $a_{1h}$ & $0.0010 \pm 0.0001$ \\
     & & $\zeta$ & $0.7 \pm 0.2$ \\
     & & $b$ & $-2$ \\
        \hline
    Blue & 2-halo & $A_\mathrm{IA}$ & $ 0.21 \pm 0.37$ \\ 
     & & $\beta$ & 0 \\
     & 1-halo & $a_{1h}$ & $0.0006 \pm 0.0002$ \\
     & & $\zeta$ & $0.5 \pm 0.4$ \\
     & & $b$ & $-2$ \\
     \hline
	\end{tabular}
\end{table}

\section{Results} \label{sec:results}

\begin{figure*}
\centering
\includegraphics[width=\textwidth]{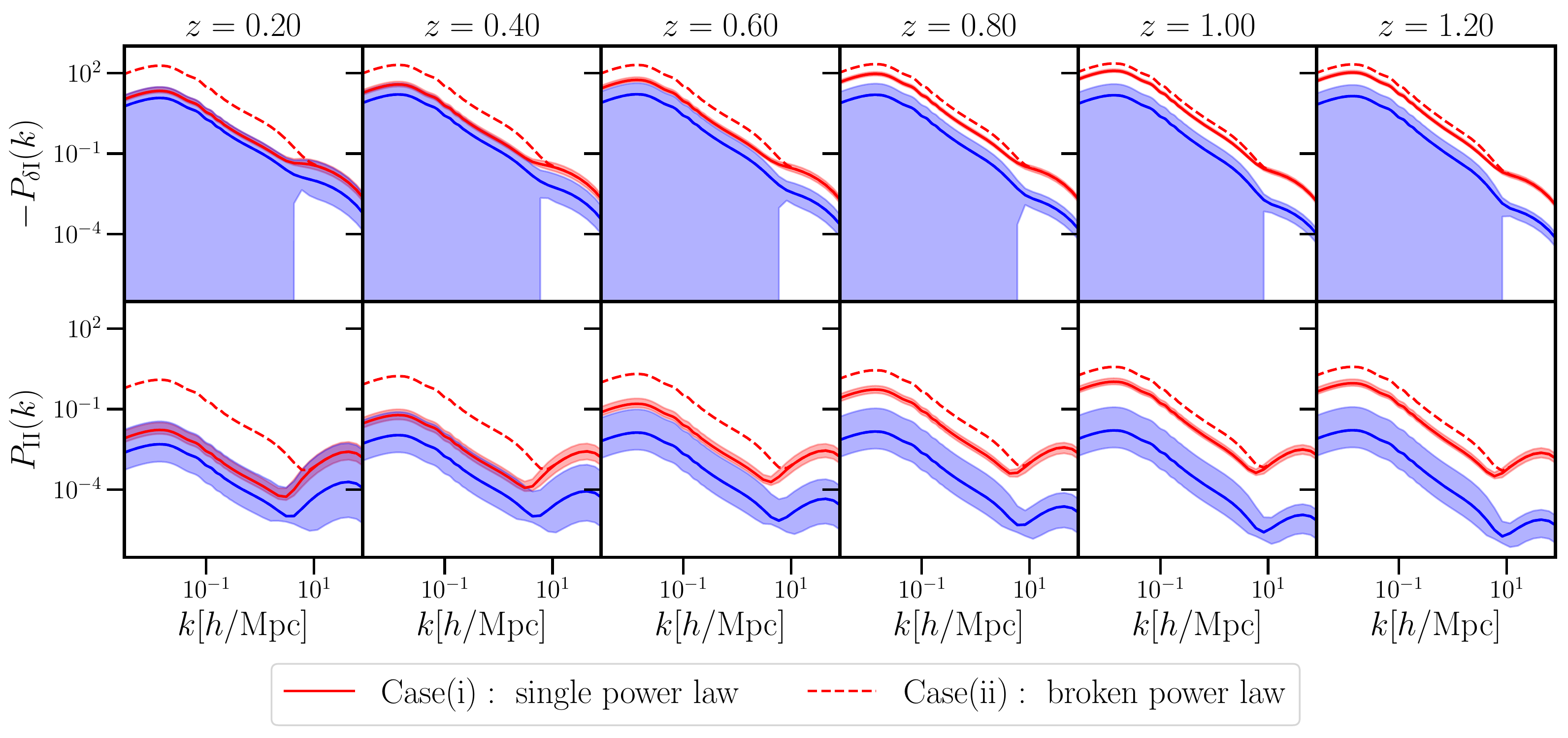}
\caption{The matter - intrinsic and the intrinsic-intrinsic shear power spectra for our case (i) (solid lines) and (ii) (dashed lines). The red curve corresponds to the red sample, the blue curve to the blue sample. The shaded areas corresponds to the $1\sigma$ uncertainties in the model as derived from the parameters uncertainties in Tab.  (\ref{tab:halo_model_setup}). For clarity, we only show it for our case (i). For the matter-intrinsic power spectra, we only plot the negative part.}
\label{fig:pks_case_i_and_ii}
\end{figure*}

Figure \ref{fig:pks_case_i_and_ii} shows our predictions for the IA power spectra for case (i) (solid lines) and (ii) (dashed lines). Our fiducial power spectra correspond to the best fit values in Table~\ref{tab:halo_model_setup}. Here, we also plot the associated $1\sigma$ uncertainties, to illustrate the current uncertainties in our IA parameters. The one-halo parameters $a_\mathrm{1h}$ and $\zeta$ are modelled as a multivariate Gaussian with the covariance matrix computed in the fitting procedure outlined in Sect.~\ref{subsec:luminosity_dependence_of_the_satellite_galaxy_alignment}. Since we do not have information on the covariance matrix between the parameters ${A_\mathrm{IA}, \beta}$, we assume that they follow uncorrelated Gaussian distributions centred on the best fit values and with standard deviation given by the $1\sigma$ uncertainties. We then draw 300 Monte Carlo realisations of the model, and we derive the lower and upper uncertainties using, respectively, the 16th and the 84th percentiles of the resulting distributions.

For both our case (i) and (ii), at low redshift and small scales, we find a larger signal that decreases as the redshifts increase, due to the drop of satellite galaxies at high redshifts imposed by the flux limit. The opposite happens for the large scales, where we observe an inverted trend in the redshift dependence: at high redshifts, where only bright galaxies are observed, the large-scale signal increases due to the luminosity dependence of the red central galaxy alignment. These trends are more pronounced for the single slope scenario (i), where we observe more variation among the different redshift bins. The power spectra of the red sample of our case (ii) maintains an almost constant effective amplitude, as expected for a luminosity distribution predominantly below $L_0$. The radial alignment of satellite galaxies shifts the contribution of the 1-halo term to larger $k$, reducing the impact of the IA at intermediate scales. 

At small scales, the uncertainty in the luminosity dependence of the blue satellite alignment dominates our predictions. We note that the joint constrains on the luminosity and radial dependence of the faint blue sample do not fully capture the measurements, as the curve always remains slightly above the data points (Fig. \ref{fig:sat_eplus_red_blue_lum_dep}). Indeed, the individual fit for the blue L2 sample is consistent with zero (Table \ref{tab:sat_lum_dep_fit}). This is driven by the fixed slope of the radial dependence when performing the combined radial and luminosity dependence fit. We find that a steeper radial dependence can capture the measurements better, but given the limited S/N we decided to not adopt separate dependencies for the red and blue samples.

\subsection{Impact on lensing} \label{subsec:impact_on_lensing}

\begin{figure*}
\centering
\includegraphics[width=\textwidth]{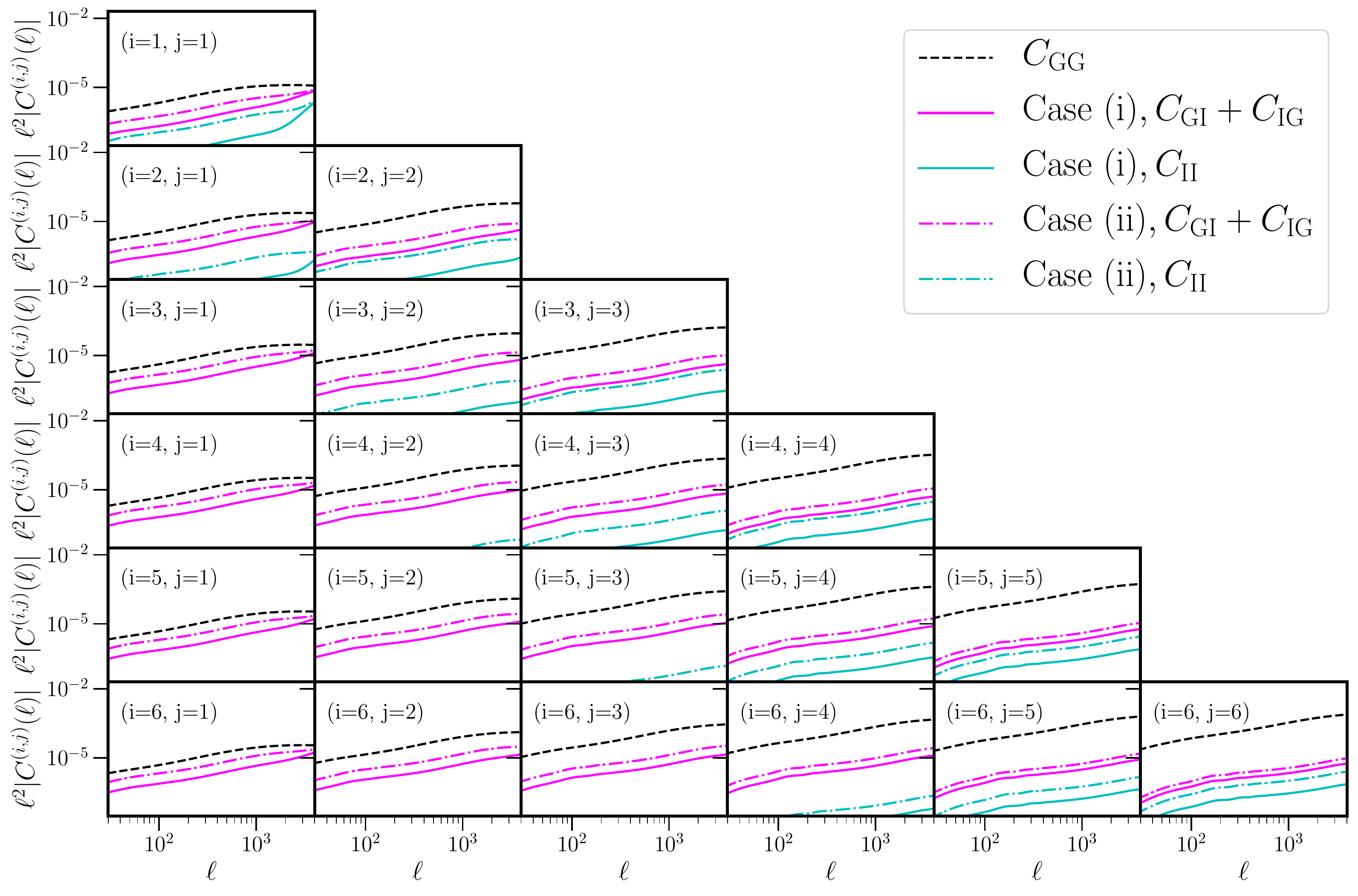}
\caption{{Predictions for the projected angular power spectra: solid lines correspond to case (i), the dash-dotted lines to case (ii). The black dashed line shows the lensing power spectrum.}}
\label{fig:cls}
\end{figure*}

To assess the contamination to the lensing measurements, we use the projected angular power spectra, $C(\ell)$, where $\ell$ is the 2D angular frequency. In the flat sky approximation, these can be written as:
\begin{equation} \label{eq:cl_obs}
    C_{\mathrm{obs}}^{(ij)}(\ell) = C_{\mathrm{GG}}^{(ij)}(\ell) + C_{\mathrm{GI}}^{(ij)}(\ell) + C_{\mathrm{II}}^{(ij)}(\ell)
\end{equation}
where
\begin{equation}
    C_{\mathrm{GG}}^{(ij)}(\ell) = \int_{0}^{\chi_\mathrm{hor}} \dd \chi \frac{q^{(i)}(\chi) q^{(j)}(\chi)}{\chi^2} P_{\delta \delta} \left(\frac{\ell}{\chi}, \chi \right),
\end{equation}
\begin{equation}
    C_{\mathrm{GI}}^{(ij)}(\ell) = \int_{0}^{\chi_\mathrm{hor}} \dd \chi \frac{q^{(i)}(\chi) p^{(j)}(\chi) + p^{(i)}(\chi) q^{(j)}(\chi)}{\chi^2} P_\mathrm{\delta I} \left(\frac{\ell}{\chi}, \chi \right),
\end{equation}
and
\begin{equation}
    C_\mathrm{II}^{(ij)}(\ell) = \int_{0}^{\chi_\mathrm{hor}} \dd \chi \frac{p^{(i)}(\chi) p^{(j)}(\chi)}{\chi^2} P_\mathrm{II} \left(\frac{\ell}{\chi}, \chi \right).
\end{equation}
Here, $\chi$ denotes the comoving distance, $\chi_\mathrm{hor}$ the comoving distance to the horizon,  $p^i(\chi) \dd \chi$ the distribution of source galaxies in the sample $i$, normalised to $\int \dd \chi \ p^i(\chi) =1 $, and $q(\chi)$ is the lensing efficiency, defined as
\begin{equation}
    q^{(i)}(\chi) = \frac{3 H_0^2 \Omega_m}{2 c^2} \int^{\chi_h}_{\chi} \dd \chi' \ p^{(i)}(\chi ') \frac{\chi'-\chi}{\chi '} \ .
\end{equation}

We compute the power spectra $P(k,z)$ for the \textit{true} redshifts and then integrate over the $n(z)$ of six photo-z bins with ranges reported in Table~\ref{tab:zbin_galinfo}. To simulate the effect of photometric scatter we generate six Gaussian redshift distributions with a scatter $\sigma_z = 0.05(1+z)$, as described in \citet{Chisar2019CCL}.  We assume a total shape dispersion of $\sigma_{\epsilon} = 0.35$. 

We compute a fully analytical covariance matrix, as described in \citet{Hildebrandt2020KV450}. To be consistent with current cosmic shear analyses, we do not include the IA contribution in the covariance matrix. To generate our predictions, we make use of the latest version of the public available software \textsc{CosmoSIS}\footnote{\url{https://bitbucket.org/joezuntz/cosmosis}} \citep{Zuntz2015CosmoSIS}. Our results are shown in Fig. \ref{fig:cls}. For clarity, here we do not show the 1$\sigma$ contours derived from the uncertainty in the IA parameters. In practise, when it comes to the contamination to lensing, the unknown luminosity dependence of the IA of faint galaxies dominates our uncertainty, being the distance between the curves labelled as (i) and (ii) larger than the individual uncertainties in the fits of the IA signal of the specific sub-samples.

At low redshift, the large fraction of satellite galaxies is reflected in the IA signal, which becomes important. As expected, the II term is only relevant in the auto-correlation bins, while the GI is larger in all of the off-diagonal terms. Overall, the relative contamination from IA is larger at low redshifts, where also lensing is less efficient.

\subsection{The impact of the modelling choice on the cosmological parameter estimate}\label{subsec:impact_of_the_modelling_choice}

The main goal of this paper is to investigate whether the emerged complexity of satellite contribution in the IA signal can lead to a bias in the Stage-III cosmological parameter estimate if not properly accounted. To explore this, we consider two cases of a generic cosmic shear analysis: in the first we simply assume the NLA model to hold for the full sample, without splitting in red and blue galaxies and without considering any luminosity dependence - so with only one free parameter, the amplitude $A_\mathrm{IA}$; in the second case we introduce a power law to capture the redshift evolution of the signal due to the IA dependence on the galaxy sample: 
\begin{equation}
    P_{\delta\mathrm{I}} (k,z) = \left( \frac{1+z}{1+z_0} \right)^{\eta} P^\mathrm{NLA}_{\delta \mathrm{I}}(k,z)
\end{equation}
and
\begin{equation}
    P_{\mathrm{II}} (k,z) = \left( \frac{1+z}{1+z_0} \right)^{2\eta} P^\mathrm{NLA}_{\mathrm{II}}(k,z),
\end{equation}
where we choose $z_0=0.3$. We refer to this model as NLA-$z$.

To do so, we generate a data vector of angular correlation functions $\xi_{\pm}(\theta)$ with the setup discussed in Sect.~\ref{sec:halo_model_setup} and analyse it assuming the NLA and NLA-$z$ as typically done in most of the Stage-III analyses. In this way, we have perfect knowledge of the signal injected and we can isolate the impact of marginalisation. 

We perform the analysis in real space, using the projected correlation functions $\xi_{\pm}$, which we derive from the angular power spectra $C(\ell)$ using the implementation available in \textsc{CosmoSIS} \citep{Kilbinger2009}. The minimum and maximum angular scales adopted in this analysis are, respectively: $\theta^\mathrm{min}_+ = 3'$, $\theta^\mathrm{max}_+ = 72'$, $\theta^\mathrm{min}_- = 6'$ and $\theta^\mathrm{max}_+ = 153'$, based on the KV450 \citep{Hildebrandt2020KV450} cosmic shear analysis.

We limit our interest to the cosmological parameters to which lensing is most sensitive, $\Omega_m$, $\sigma_8$ and $w$. Instead of $\sigma_8$, we sample the logarithm of the scalar amplitude $\ln (10^{10} A_s) $, so our final parameter vector is $\lambda = \{ \Omega_m, \ln (10^{10} A_s), w \}$ and one (two) nuisance parameter(s), $A_\mathrm{IA}$ ($A_\mathrm{IA}$, $\eta$). We adopt uniform priors $\Omega_m = \left[0.1,0.8 \right]$, $\ln (10^{10}A_s ) = \left[ 1.5, 5\right]$, $w \in [-5.0, 0.33]$, $A_\mathrm{IA} = \left[-6,6 \right]$ ($\eta = [-5,5]$). To sample the parameter space we make use of the \textsc{Emcee} sampler \citep{Foreman-Mackey2013emcee}. The same analysis is performed for both scenarios; we only change the IA recipe while generating the data vector.

Our results show that for Stage-III surveys the NLA model provides an adequate description. For both scenarios the redshift dependence of the IA signal caused by the variation of the galaxy sample is not large enough to induce a bias in the cosmological parameters, with only marginal shifts in both $S_8$ and $w$ for our case (ii). The recovered IA amplitudes are instead, as expected, different. In our case (i) we find a $A_\mathrm{IA, (i)} = 0.14\pm 0.14$, while for our case (ii) we find $A_\mathrm{IA, (ii)} = 0.44\pm 0.13$.

When adopting the NLA$-z$ model as the reference, in both cases the cosmological parameters are correctly recovered, but the $\eta$ parameter remains unconstrained in our case (i) and it is very weakly constrained in our case (ii).

Our low IA amplitude for case (i) is in line with the best fit NLA amplitude found in \citetalias{Johnston2019} for the full GAMA sample, while their best fit value for the joint GAMA+SDSS Main has an amplitude of $A_\mathrm{IA} \sim 1$, compatible with the fact that SDSS Main contains a larger fraction of red galaxies and fewer satellites to lower the signal at large scales. The comparison is, however, complicated by the fact that \citetalias{Johnston2019}'s results are based on the gI correlations. Compared to the KV450 IA amplitude, $A_\mathrm{IA} = 0.981^{+0.694}_{-0.678}$, we find our case (i) to provide a lower value for a similar galaxy sample. However, different redshift distributions are adopted in the two works. We note that the full shape of the $n(z)$ is critical for the accurate modelling of the IA contribution (see Appendix \ref{A:ia_dependence_on_nz}). The redshift distributions of KV450 are more peaked and with more prominent tails, which increase the impact of the II in real data: as a consequence, since II and GI have opposite contributions, the IA balance changes. The way this effect can couple with the IA sample dependence is not obvious, as calibration errors in the final $n(z)$ can be absorbed by the $A_\mathrm{IA}$ amplitude during the fit \citep[][]{Li2020KV450colorsplit}. In addition to this, a luminosity dependence of the signal reduces the presence of IA in the data \citep{Joachimi2011b, Krause2016}: if the faint end of the luminosity dependence of red central galaxies is significantly shallower than what we assumed in our case (i), the final amplitude would increase, as already suggested by our case (ii) setup. Similarly, our predictions are based on the assumption that the blue central galaxy population does not significantly contribute to the signal ($A^\mathrm{blue}_\mathrm{IA} = 0.21$), a constraint that suffers from large uncertainties. Our results (i) and (ii) point toward a lower amplitude to what is preferred by the cosmic shear analysis in DES data, for both their results with the NLA and NLA$-z$ models \citep[][]{Troxel2018DESY1}, although we observe the same increase of the overall IA amplitude as a function of redshift. \citet[][]{Samuroff2018DES} find a lower IA amplitude in DES galaxies when simultaneously fitting for the cosmology and the IA amplitude in a 3x2pt statistics ($\gamma \gamma + \delta_g \gamma + \delta_g \delta_g$, $A_\mathrm{NLA} = 0.49^{+0.15}_{-0.15}$), which is in closer agreement with our findings. We want to stress that the aim of these comparisons is only to provide a sense of the ranges of the IA amplitudes currently constrained by lensing analyses: we should not interpret the IA amplitudes as stand-alone quantities, without taking into account the best fit cosmological parameters and the exact $n(z)$.

\subsubsection{Stage-IV}

\begin{figure}
\centering
\includegraphics[width=\columnwidth]{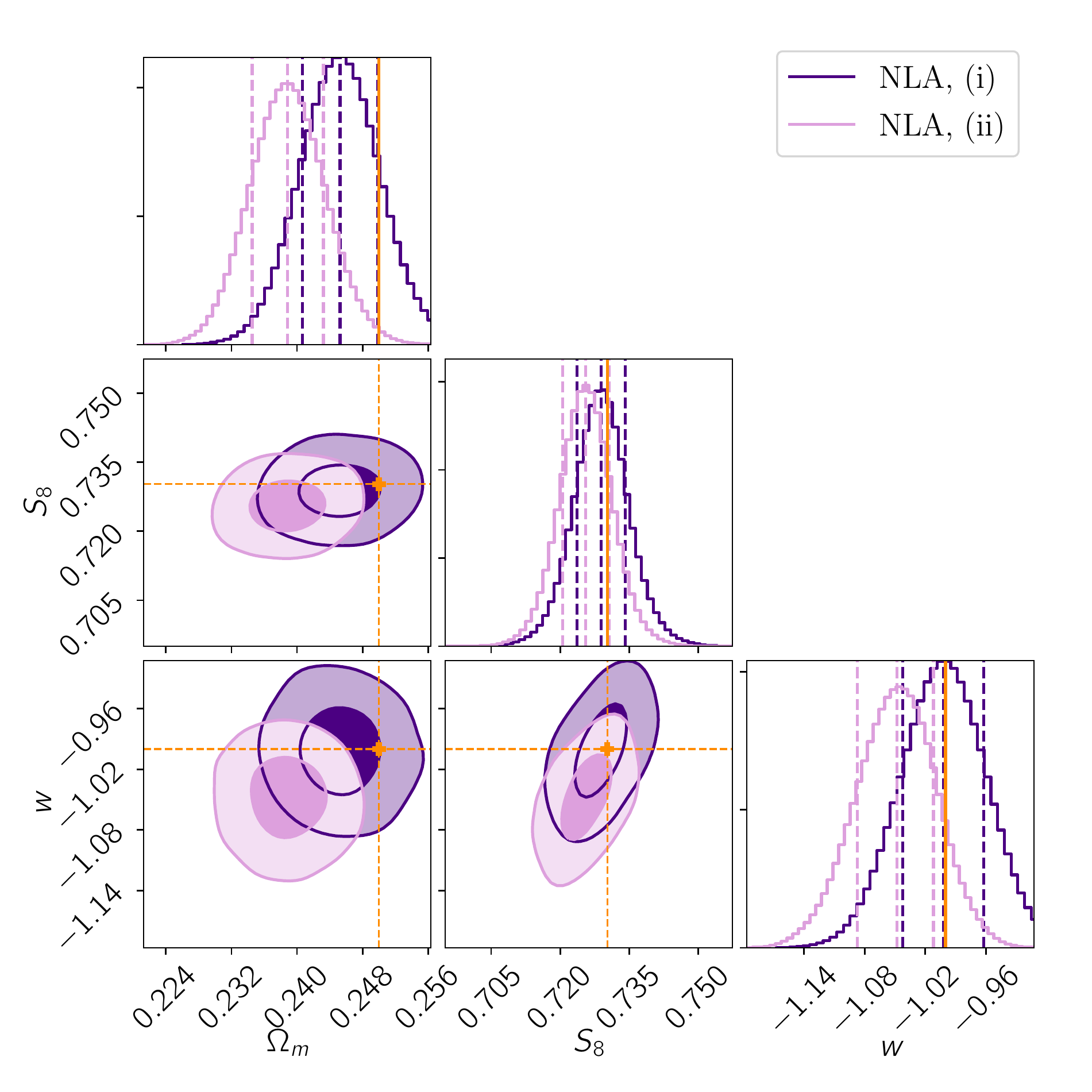}
\caption{Constraints on the cosmological parameters $\Omega_m$, $S_8 = \sigma_8 \sqrt{\Omega_m/0.3}$ and $w$, marginalising over the IA amplitude, for a Stage-IV survey. We inject the IA signal as predicted by the full halo model formalism assuming at larger scales a steep luminosity dependence (i: indigo) or a broken power law with constant amplitude for faint galaxies (ii: plum) - Table~\ref{tab:halo_model_setup} lists the IA parameters used for constructing the data vectors. We perform the analysis assuming a NLA model with no distinction between red and blue galaxies. The orange lines and the square markers indicate the fiducial values of the cosmological parameters.}
\label{fig:corner_NLA}
\end{figure}

\begin{figure}
\centering
\includegraphics[width=\columnwidth]{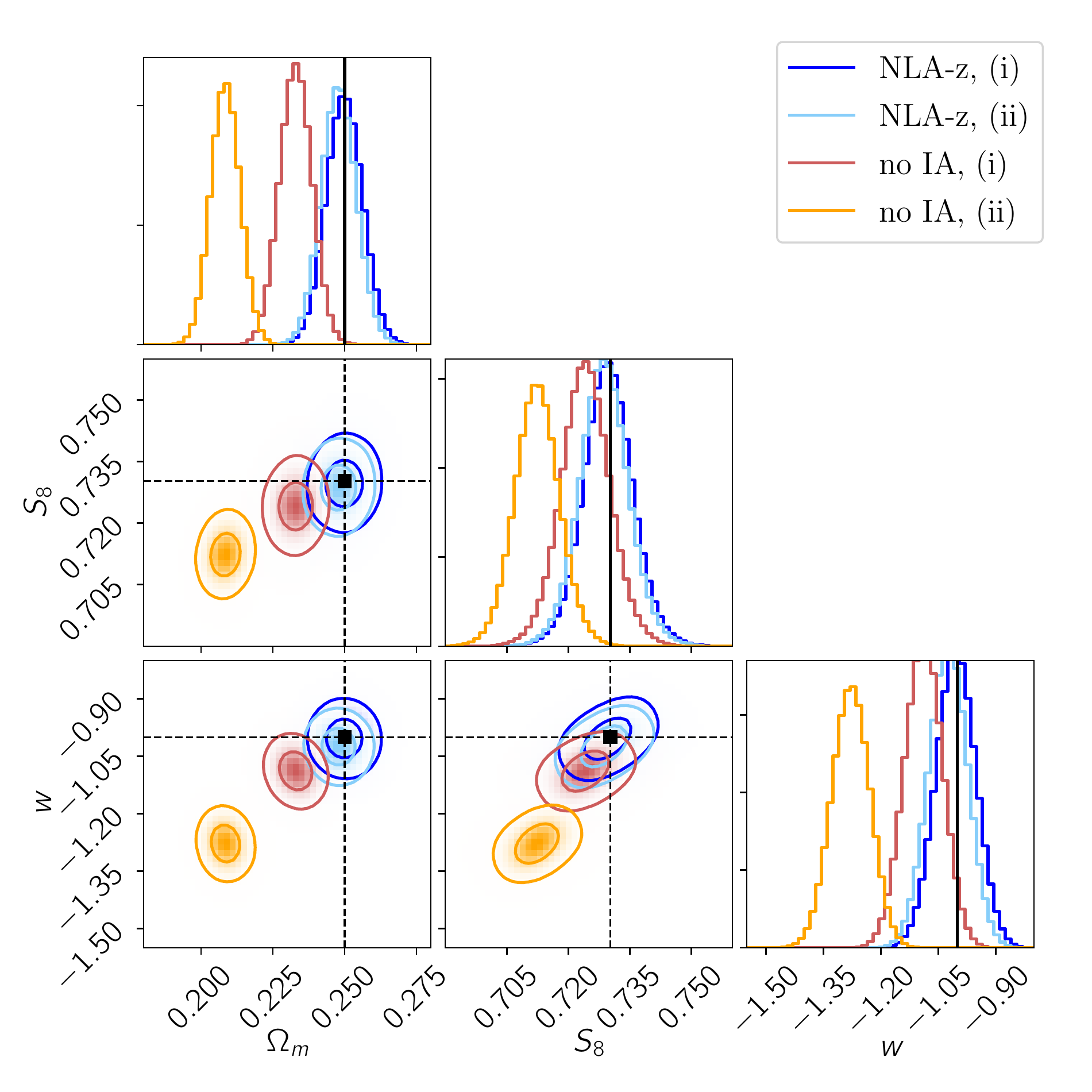}
\caption{Constraints on the cosmological parameters $\Omega_m$, $S_8 = \sigma_8 \sqrt{\Omega_m/0.3}$ and $w$, marginalising over the IA amplitude, for a Stage-IV survey. We distinguish between two IA scenarios at large scales, as detailed in Table \ref{tab:halo_model_setup}. We perform the analysis assuming no distinction between red and blue galaxies and assume either a NLA$-z$ model (dark blue ellipse for the IA signal as in case (i) and light blue ellipse for the case (ii)) or no IA (red for the IA signal based on (i), orange for the (ii)). The black lines and the square markers indicate the fiducial values of the cosmological parameters.}
\label{fig:corner_NLA}
\end{figure}

Given our results on a Stage-III setup, we investigate whether in the case of a Stage-IV survey we still recover the right cosmological parameters. We leave our setup unchanged, and only replace the covariance matrix to account for the larger area (15 000 deg$^2$) and double the number density per redshift bin. We note that we do not modify our HODs, resulting in an underestimate of the satellite number density. Thus, our results have to be considered a lower limit on the possible induced biases. 

Fig. \ref{fig:corner_NLA} illustrates our findings for the NLA model. Overall, we find that case (i) leads to a lower level of bias in all the cosmological parameters compared to case (ii); this is expected, given the lower IA signal present in the data when assuming a steep luminosity dependence. In our case (i) we observe a $1\sigma$ bias in $\Omega_m$ only, while for case (ii) all parameters are biased by
more than $1\sigma$, with the bias in $\Omega_m$ exceeding $2\sigma$. We note that the NLA $A_\mathrm{IA}$ amplitude recovered by the fits is dominated by the effective amplitude of the low redshift bins. The relative importance of IA over lensing is indeed strongest when the foreground galaxies are at low redshift (see the first columns of Fig. \ref{fig:cls}) and thus, even if the data contain a significantly higher alignment signal at high redshift, the fit is not particularly sensitive to these bins.

We also explore the performance of the NLA$-z$ model for the same setups. We find the flexibility of the NLA$-z$ sufficient to recover the cosmological parameters for both our case (i) and (ii), with best-fit values: $A_\mathrm{IA,(i)} = 0.16^{+0.02}_{-0.02}$, $\eta_\mathrm{(i)} = 2.91^{+0.68}_{-0.73}$ and $A_\mathrm{IA,(ii)} = 0.42^{+0.02}_{-0.02}$, $\eta_\mathrm{(ii)} = 2.21^{+0.22}_{-0.23}$. The predicted redshift dependence of the signal for our sample can be captured by its power-law scaling, while the scale dependence introduced by the 1-halo term is not recovered. Because the amplitude of the 1-halo term is small and we remove the smallest scales in our fit with the cuts presented in Sec. \ref{subsec:impact_of_the_modelling_choice}, this remains a subdominant effect compared to the redshift variation in the IA signal induced by the galaxy sample selection across the redshift tomographic bins. In this case, we also note that the fits are driven by the low-z bins, with a worse recovery of the large scale alignment at high redshifts. This is more pronounced for case (ii), where the double power-law induces a more complex redshift scaling in the IA signal.

Given the precision of a Stage-IV survey, any specific choice in the setup can impact the results. We identify as the most significant ingredient the treatment of the halo exclusion. We caution that our implementation of the halo exclusion is not based on simulations, but merely aims to avoid the double counting at small scales, due to the use of the NLA$-\beta$ model for the 2-halo term. While this is sufficient for a Stage-III survey, we observe that the bias in the cosmological parameters is affected by the specific implementation of the halo exclusion for a Stage-IV. In general, a smoother transition that would remove more power at the intermediate scales would result in less bias in $\Omega_m$ and more bias in $S_8$ and $w$. The IA amplitude is instead weakly affected by the specific choice of the halo exclusion recipe. We also note that a full double counting at the small scales does not significantly change our results. We conclude that a proper modelling of the intermediate scales is more important than the exact amplitude of the 1-halo term.

\section{Conclusions}\label{sec:conclusions}

We have performed a comprehensive analysis of the contamination by IA in cosmic shear surveys, with a particular focus on modelling the satellite contributions at small and large scales, based on the most recent observational IA findings. We proposed a new model to describe the IA signal, which explicitly accounts for the fact that only central galaxies contribute to the alignment signal at large scales. We introduced a satellite alignment signal at small scales, modelled through the halo model formalism, which includes a radial and luminosity dependence. We also differentiated the contribution from the red and blue population at all scales.

At large scales, we investigated whether limiting a luminosity dependence of the IA signal to the central galaxy sample provides a unified picture for all the measurements in the literature. Although in this scenario the slope measured by MegaZ/LOWZ seems to be favoured at high luminosities, the current uncertainty in the measurements does not allow for a definitive constraint on the luminosity dependence at low luminosities. For this reason, we decided to follow two alternative scenarios for all our forecasts: a single power law with slope $\beta=1.2$ and a double power law with a faint end characterised by a flat slope, $\beta_{L<L_0}=0$, and a bright-end with same slope as case (i), $\beta_{L \ge L_0}=1.2$.  Future IA studies should focus on constraining the faint end of Fig. \ref{fig:A_L_gama_cencen}, where uncertainties dominate. Upcoming surveys such as the Physics of the Accelerating Universe Survey\footnote{\url{https://www.pausurvey.org}} \citep[PAUS;][]{Eriksen2019,Padilla2019} can help gaining insight into our understanding of the faint central galaxy alignment, a key feature to properly predict the IA contamination in cosmic shear surveys. 

At small scales, we model the satellite alignment with a power law for the radial dependence, as recently measured by \citetalias{Georgiou2019b} in groups. We re-analyse the \citetalias{Georgiou2019b} data splitting the sample in red and blue, and we found that in this case a cut at $M_r=-22$ suggests a luminosity dependence in the signal. We jointly fitted the radial and luminosity dependence assuming a double power law, and used this result as input for our forecasting model. More data are needed to tightly constrain the luminosity dependence of satellites, as the statistical uncertainties in the measurements might play a role in constraining the amplitude of the power law. It is also relevant to note that since red and blue satellites show a different alignment amplitude, it is important to model them separately, as their relative fraction depends on the magnitude cut of the specific survey.

Although satellites do not share the same alignment mechanisms as central galaxies, the dichotomy in morphology observed at large scales is reflected also in their alignment mechanism. A different radial dependence of red and blue satellites might reveal a more complex alignment mechanism for the two populations and/or probe the galaxy in-fall history. With current measurements it is not possible to further investigate this possibility, but future dedicated high resolution hydrodynamical simulations might shed light in the understanding of the intra-halo alignment. Such simulations can also improve the modelling of intermediate scales, to which the lensing signal is most sensitive, but where our model is relatively simplistic. A proper calibration of the IA alignment in this intermediate regime is of primary importance for the interpretation of data from Stage-IV surveys.

Our predicted power spectra show two opposite trends at high and low redshifts, as shown in Fig. \ref{fig:pks_case_i_and_ii}. While at low redshift the small scales have a larger IA signal, due to the presence of satellites, the large scales are dominated by the alignment of red central galaxies. The large scale signal becomes stronger as we go to higher redshifts, due to the survey magnitude cut: this selects brighter galaxies, which are those that carry most of the IA signal, due to the observed luminosity dependence of the red central galaxy alignment. The opposite happens to the small scales, which are almost completely washed out by the suppression of satellites in the high redshift bins. These trends are enhanced when considering the single power law for the luminosity dependence.

In this work we have not accounted for the anisotropic distribution of satellites within the halo, which is also known to contaminate lensing measurements. In the context of cosmic shear analyses, this is expected to be important only at small and intermediate scales. The satellite segregation along the central galaxy major axis complicates the interpretation of IA measurements performed using the central galaxy position - satellite shear correlation (gI), and future studies should focus on modelling it, in order to have a clear mapping between gI and GI. However, our spherical model can be considered as an \textit{effective} model: in the perspective of a direct fit of IA to data, the free amplitude in the 1-halo term can potentially capture to first order the extra correlation due to the anisotropic term.

While direct IA measurements provide unique insights into the IA mechanisms and amplitudes, translating those results into informative priors for cosmic shear analyses requires a full modelling of the sample dependence of the IA signal. This aspect has often been underestimated in lensing studies, adopting simplistic models that do not distinguish between the different IA signatures of different galaxy populations, and adopting broad, uninformative priors. We investigated what is the impact of this choice for the IA signal that we expect for a cosmic shear survey given our model. We considered the case of a Stage-III analysis on a simulated data vector, built to reproduce our best knowledge of the IA, and then analysed it using simple NLA and NLA$-z$ models. Limiting our analysis to $\Omega_m, S_8$ and $w$, we find both models to be sufficient to capture the IA signal without biasing in the cosmological parameters. 
This is no longer true for Stage-IV surveys, where we observe 
a bias in $\Omega_m$ that exceeds $2\sigma$ bias when adopting a simple NLA model. Including a power-law redshift dependence, the NLA$-z$ model
is able to recover our input cosmological parameters in the presence of a sample-dependent IA signal. Hence we recommend the use of a flexible redshift dependent model of IA for Stage-IV surveys.  

In all cases the recovered IA amplitudes are smaller than 1, similar to typical values of obtained by current cosmic shear studies. The amplitude depends on the specific IA model assumed: we find the broken power law for the red central galaxy luminosity dependence to provide a larger effective amplitude. This is a consequence of the larger amplitude assumed for the faint population in this setup, which dominates our overall alignment, as faint galaxies are more abundant than bright galaxies. We find the IA signal to be smaller than what was assumed in previous works: this is driven by the central galaxy weighting at large scales which significantly reduces the effective IA amplitude. This has implications for the inferred level of bias in the cosmological parameters. These findings are, however, based on several assumptions that represent our best extrapolation of the current picture of IA as emerging from dedicated studies. Future studies on the behaviour of faint central and satellite galaxies are needed to confirm these results. However, the model is extremely flexible and any new findings can be easily incorporated. 

We find our fits to be driven by the 
low-z bins, where IA dominates over cosmic shear, whereas at high-z a bias in IA barely affects the results. We also note that the redshift dependence of the signal is more important than the scale dependence introduced by the 1-halo term. This is a consequence of the small satellite alignment we adopt in the model, based on current observational constraints. The inclusion of a term for the anisotropic distribution of the satellites might change this conclusion.

As the impact of IA is larger for the lowest redshift bin, excluding it can mitigate the impact of uncertainties in the modelling of the IA signal. Alternatively, improving the constraints at low-$z$, with a focus on faint galaxies, will be essential.

We want to stress that these model predictions are based on idealised redshift distributions and so our results cannot be directly compared with the best fit parameters inferred by cosmic shear studies. At higher redshifts and lower luminosities, the mocks might suffer from a larger uncertainty, as the luminosity is calibrated locally and then evolved to high redshifts. This has to be considered as part of the uncertainty in the model predictions, which requires additional data to be assessed. Upcoming surveys that aim to use the model can use clustering information in their data and the observed luminosity function to further constrain these parameters.

In light of future cosmological analyses, this model can be used to account for the IA signatures on different galaxy populations allowing for a direct map between IA observations and cosmic shear contamination. The model can be used to predict the IA signal in a given cosmic shear based on its galaxy composition at different redshifts: provided that the galaxy mocks are sufficiently representative of the data, it is possible to provide priors for different IA models. We caution that the current uncertainty in the luminosity dependence of the signal currently prevents the use of tight priors; instead the range covered by both out (i) and (ii) scenarios should be considered. Given a halo occupation distribution model for the red and blue galaxy populations, this model can be employed to jointly fit the clustering, IA and lensing observables.

\section*{Acknowledgements}

We thank Shahab Joudaki and Francisco Castander for useful discussions and comments and Kai Hoffmann for providing comments on a preliminary version of this manuscript. We thank Jonathan Blazek, the referee, for his comments, which have greatly improved the final manuscript. MCF and HH acknowledge support from Vici grant 639.043.512, financed by the Netherlands Organisation for Scientific Research (NWO). HH also acknowledges funding from the EU Horizon 2020 research and innovation programme under grant agreement 776247.

The MICE simulations have been developed at the MareNostrum supercomputer (BSC-CNS) thanks to grants AECT-2006-2-0011 through AECT-2015-1-0013. Data products have been stored at the Port d'Informaci$\acute{\mathrm{o}}$ Cient$\acute{\mathrm{i}}$fica (PIC), and distributed through the CosmoHub webportal (cosmohub.pic.es). Funding for this project was partially provided by the Spanish Ministerio de Ciencia e Innovacion (MICINN), projects 200850I176, AYA2009-13936, AYA2012-39620, AYA2013-44327, ESP2013-48274, ESP2014-58384 , Consolider-Ingenio CSD2007- 00060, research project 2009-SGR-1398 from Generalitat de Catalunya, and the Ramon y Cajal MICINN program.

Based on data products from observations made with ESO Telescopes at the La Silla Paranal Observatory under programme IDs 177.A- 3016, 177.A-3017 and 177.A-3018, and on data products produced by Tar- get/OmegaCEN, INAF-OACN, INAF-OAPD and the KiDS production team, on behalf of the KiDS consortium. OmegaCEN and the KiDS production team acknowledge support by NOVA and NWO-M grants. Members of INAF-OAPD and INAF-OACN also acknowledge the support from the Department of Physics $\&$ Astronomy of the University of Padova, and of the Department of Physics of Univ. Federico II (Naples). GAMA is a joint European-Australasian project based around a spectroscopic campaign using the Anglo-Australian Telescope. The GAMA input catalogue is based on data taken from the Sloan Digital Sky Survey and the UKIRT Infrared Deep Sky Survey. Complementary imaging of the GAMA regions is being obtained by a number of independent survey programmes including GALEX MIS, VST KiDS, VISTA VIKING, WISE, Herschel-ATLAS, GMRT and ASKAP providing UV to radio coverage. GAMA is funded by the STFC (UK), the ARC (Australia), the AAO, and the participating institutions. The GAMA website is \url{http://www.gama-survey.org/}.

\section*{Data availability}
This work has made use of the MICECATv2 simulation, publicly available here: \url{http://maia.ice.cat/mice/}. The other data underlying this article will be shared on reasonable request to the corresponding author.




\bibliographystyle{mnras}
\bibliography{iabiblio} 




\appendix

\section{Satellite galaxy fractions in MICE}
\label{A:satellite_galaxy_fractions_in_mice}

\begin{figure}
\centering
\includegraphics[width=\columnwidth]{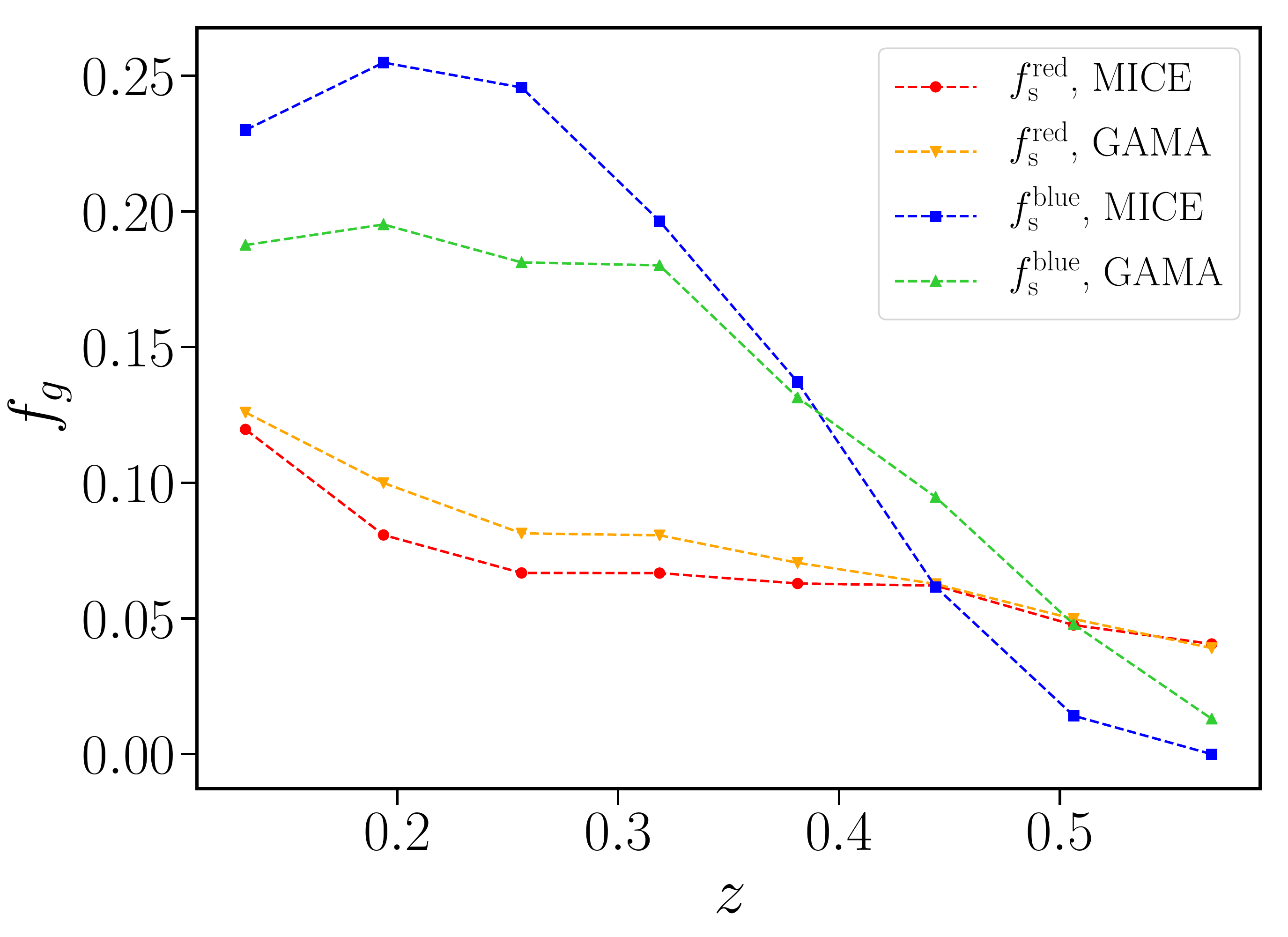}
\caption{The satellite galaxy fractions in our mocks compared to those in the GAMA sample. We select the galaxies in the mocks to reproduce the same redshift-magnitude selection as GAMA.}
\label{fig:galaxy_fractions_gama_comparison}
\end{figure}
We investigate here how well MICE reproduces the GAMA survey in the corresponding redshift and magnitude space, because we use that sample as input for our study. Our reference GAMA catalogue is obtained by matching the StellarMassLambda-v2 catalogue, from which we obtain photometric information, and the G3CGal, which contains the group information. We identify the satellite population by imposing RankBCG>1. 

We build a GAMA-like sample from MICE in the following way. We select from our mock all galaxies brighter that $r<19.8$, the flux-limit of our GAMA catalogue, and limited the analysis to a sub-patch of $\sim 180$ deg$^2$, which is the size of the overlapping region with the KiDS survey employed by \citetalias{Johnston2019}, C19 and in this work. With only these conditions, we find a very good agreement with the colour and magnitude distributions of the GAMA galaxies.

Given the compatibility in the colour-magnitude space, we select red galaxies in GAMA by applying the same cut as used in the rest of the paper (\ref{subsec:galaxy_mocks}). We then measure the galaxy fraction for all the sub-samples that enter in the halo model, to quantify the accuracy of our mocks. Fig. \ref{fig:galaxy_fractions_gama_comparison} illustrates our findings. We observe a remarkable agreement for the red galaxy population. For the blue sample, MICE exhibits a larger satellite fraction at low redshift and the opposite behaviour at high redshifts; however, the absolute difference between the fractions found in MICE and GAMA is 0.065 at maximum.

\section{Halo exclusion and intermediate scales} \label{A:halo_exclusion}

One of the limitations of the halo model is its lack of a proper treatment of the intermediate, mildly non-linear scales. In particular, the configuration with radially dependent satellite alignment pushes the one-halo term to very small scales ( $k<6 h/\mathrm{Mpc}$). This implies that the intermediate scales do not arise as a simple sum of the 1h and 2h terms, with the one halo term being too small in this regime. Current observations of the IA signal show a much stronger alignment at intermediate scales than what is predicted by this model. However, a further complication comes from the fact that the intermediate scales are significantly affected by the central-satellite correlation, which is known to be large in the context of gI, due to the anisotropic distribution of satellites within the halo. Since direct IA measurements are performed correlating galaxy positions with galaxy shapes, we lack a proper reference for this term in the case of the matter-shear and shear-shear correlations. A proper treatment of this problem requires a dedicated calibration with simulations, which we defer to future work. 

To compensate the lack of power at intermediate scales, we use the non-linear power spectrum as done in the context of the NLA model. This leads, however, to double counting at the level of the 1-halo term. To avoid this, we truncate the 2-halo term at $k_\mathrm{2h}=6 h/$Mpc, roughly corresponding to a halo of $1 \ \mathrm{Mpc}/h$, via a window function of the form:
\begin{equation}
    f^\mathrm{2h-trunc}(k) = \exp \left[- \left(k/k_\mathrm{2h}\right)^2 \right] \ .
\end{equation}

Similarly, we truncate the 1-halo term to  $k_\mathrm{1h}=4 h/$Mpc applying
\begin{equation}
    f^\mathrm{1h-trunc}(k) = 1-\exp \left[- \left(k/k_\mathrm{1h} \right)^2 \right] \ .
\end{equation}

We allow for a small overlap of the signal at intermediate scales, where the aforementioned truncations gradually reduce both the 1h and 2h terms.

\section{The angular part of the satellite alignment density run}
\label{A:angular_terms_in_IA}

We compute the satellite alignment following the formalism developed in \citet{SchneiderBridle2010}, assuming a perfect radial alignment scenario. 
We report here our expansion for $f_l$ and the main steps to derive it. We have tested our results against the analytical solution for the first two non-zero multipoles ($l_\mathrm{max}=4$), finding excellent agreement.

The complex phase in equation \ref{eq:gamma_fourier_transform} can be re-written through the plane wave expansion:
\begin{equation}
    \exp(i \mathbf{k} \cdot \mathbf{r}) = \sum_{l=0}^{\infty} i^l (2l+1) P_l(\cos \gamma) j_l(kr) \ ,
\end{equation}
where $P_l(x)$ are the Legendre polynomials of order $l$ and $\cos \gamma$ is the angle between $\mathbf{r}=(r, \theta, \phi)$ and $\mathbf{k}=(k, \theta_k, \phi_k)$,
\begin{equation}
    \label{eq:cos_gamma}
    \cos \gamma = \sin \theta_k \sin \theta \cos \left( \phi_k - \phi \right) + \cos \theta_k \cos \theta \ .
\end{equation}
We can rewrite the Legendre polynomials through the identity
\begin{equation}
    P_l (x) = 2^l \sum_{m=0}^{l} x^m \binom{l}{m} \binom{\frac{l+m-1}{2}}{l} \ ,
\end{equation}
where $x=\cos \gamma$. We can express $(\cos \gamma)^m$ using the binomial theorem, such that all of the terms on the right in equation \ref{eq:cos_gamma} are of the form $a^{j}$ and those on the left $b^{m-j}$, where $j$ goes from 0 to $m$. Separating the integrals on $\phi$ and $\theta$, we have that the azimuthal part reduces to
\begin{equation}
    \int_0^{2\pi} \dd \phi \binom{\cos (2 \phi) }{\sin (2 \phi)} \left( \cos \phi_k \cos \phi + \sin \phi_k \sin \phi \right)^j \ , 
\end{equation}
where the integrals over $\phi$ are all zeros for odd $j$, while the even terms are
\begin{equation}\label{eq:g_j}
    g_{j=2n} = \left\{ 0, \frac{\pi}{2}, \frac{\pi}{2}, \frac{15\pi}{32}, ... \right\} \mbox{for n=1,2,3, \dots} \ .
\end{equation}
The terms in $\phi_k$ can be rearranged to be of the form $\mathrm{e}^{i 2\phi_k}$. 
The integral over the polar angle $\theta$ gives instead
\begin{equation}
\begin{aligned}
    \int_{-1}^{1} \dd \cos \theta \sin^{j+1} \theta &\cos^{m-j} \theta  = \int_{-1}^{1} \dd \cos \theta \left( 1 - \cos^2 \theta \right)^{\frac{j+1}{2}}  \cos^{m-j} \theta \\ &= I(j+1,m-j) \ ,
\end{aligned}
\end{equation}
where, following \citetalias{SchneiderBridle2010}, we have defined
\begin{equation}
  I(a,b) = \int_{-1}^{1} \dd x (1-x^2)^\frac{a}{2}x^b  .
\end{equation}

Collecting all the terms together, we get
\begin{equation}\label{eq:fell_my_version}
\begin{aligned}
    f_l(\theta_k, \phi_k) = \mathrm{e}^{i 2 \phi_k} 2^l &\sum_{m=0}^{l} \binom{l}{m} \binom{\frac{l+m-1}{2}}{l} \sum_{j=0}^m \binom{m}{j} g_j \\ &\times I(j+1,m-j) \sin^j \theta_k \cos^{m-j} \theta_k \ .
\end{aligned}
\end{equation}

\begin{figure}
\centering
\includegraphics[width=\columnwidth]{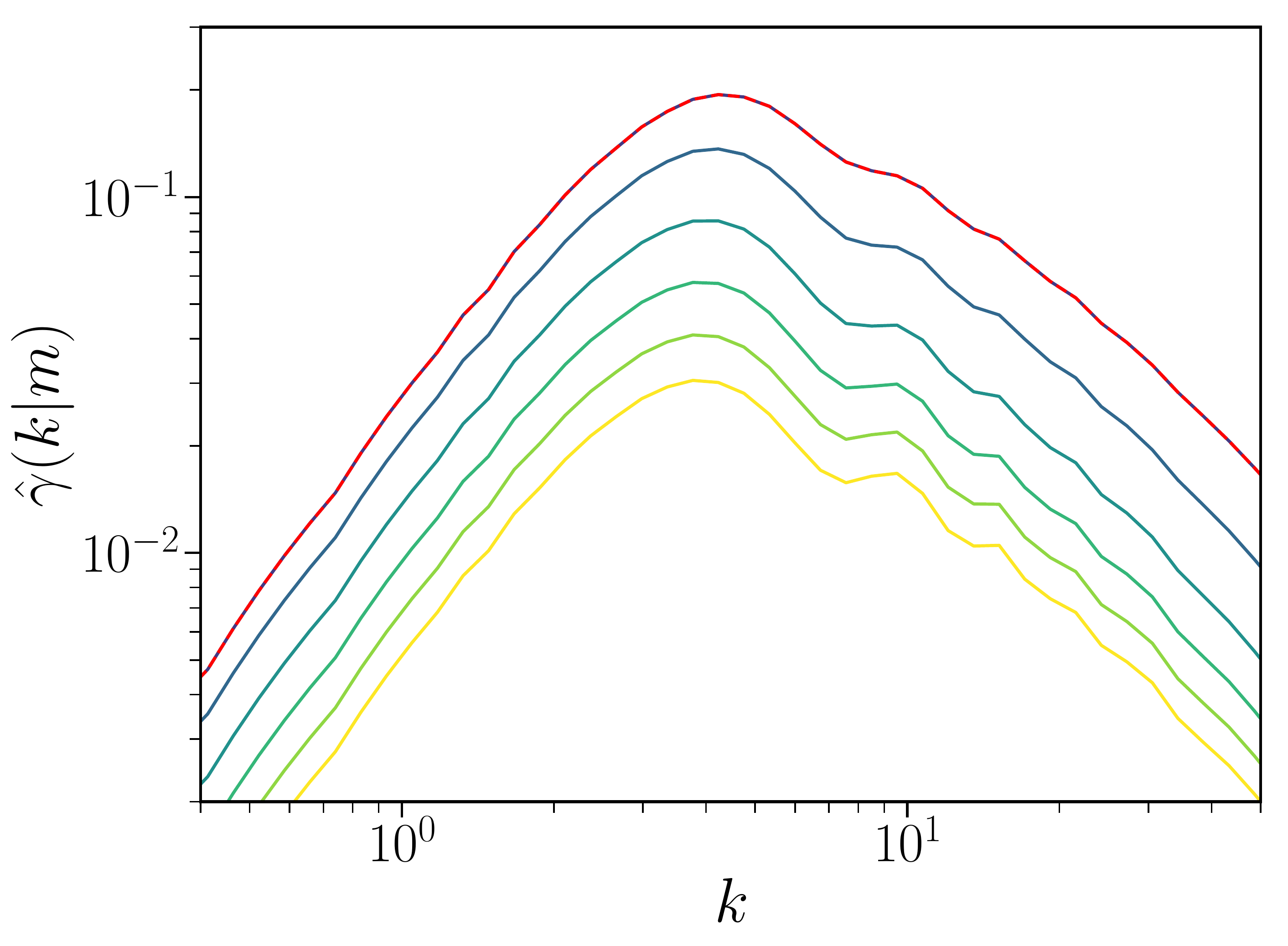}
\caption{The Fourier transform of the density-weighted shear (eq. \ref{eq:gamma_fourier_transform}) for $\phi_k =0$ and $\theta_k \in \left[\frac{\pi}{2}, \pi \right]$, in the case of constant radial dependence. For clarity, we normalise the curves by the input amplitude, $a_\mathrm{1h}$. The amplitude of the curves decreases as we go from $\theta_k=\frac{\pi}{2}$ (red) to $\theta_k=\pi$. Note that $f(\theta_k)$ is symmetric around $\theta_k=\pi/2$, so the curves from $\left[0,\frac{\pi}{2} \right]$ coincide with the ones plotted here, with increasing amplitude for increasing values of $\theta_k$.}
\label{fig:wkm_theta_k}
\end{figure}

Since the $E-$ and $B-$ modes of the intrinsic alignment are invariant under rotation in the plane of the sky, we can choose without loss of generality to fix $\phi_k=0$. The polar angle $\theta_k$ defines the projection of the wave vector $\mathbf{k}$ on the plane of the sky: modes perpendicular to the line of sight are identified by $\theta_k=\frac{\pi}{2}$, for which we have the strongest alignment signal, as illustrated in Fig. \ref{fig:wkm_theta_k}. Indeed, the angular part of $\hat{\gamma}(\mathbf{k},M)$ is dominated by the lowest term of the expansion, $l=2$, which peaks at $\theta_k=\pi/2$. The main effect of $\theta_k$ is to change the amplitude of $\hat{\gamma}(\mathbf{k},M)$ (Fig. \ref{fig:wkm_theta_k}). We decide to assume $\theta_k=\frac{\pi}{2}$ throughout our analysis, i.e. to only consider the modes perpendicular to the line of sight, and to truncate the expansion at $l_\mathrm{max}=6$.

Note that we decide to not adopt the definition of the density-weighted shear in \citetalias{SchneiderBridle2010}, $w(k|M)$ and instead work with their original definition (their equation 7), from which we can naturally derive the expression for the radially dependent case. Here, we normalise the density-weighted shear with the NFW mass, as originally in eq. 7 of \citetalias{SchneiderBridle2010}.

In our work, the intrinsic shear has the form 
\begin{equation}
    \gamma^I(r, \theta, L) \equiv \bar{\gamma}(r,L) \sin \theta = a_\mathrm{1h}(L) \sin^b \theta \left( \frac{r}{r_\mathrm{vir}} \right)^b \ ,
\end{equation}
where the $\sin^b \theta$ in the right-hand side comes from the de-projection of the satellite separation from the BCG. 
This brings $\sin \theta \mapsto \sin^b \theta$ and only affects the integral in $\dd \cos \theta$, such that $I(j+1,m-j) \mapsto I(j+b,m-j)$ in eq.~\ref{eq:fell_my_version}. To avoid singularities along the line-of-sight, we perform the integral in the range $[-1+\varepsilon, 1-\varepsilon]$, with $\varepsilon=10^{-10}$.

\section{Intrinsic alignment dependence on photometric redshift distributions}\label{A:ia_dependence_on_nz}

\begin{figure}
\centering
\includegraphics[width=\columnwidth]{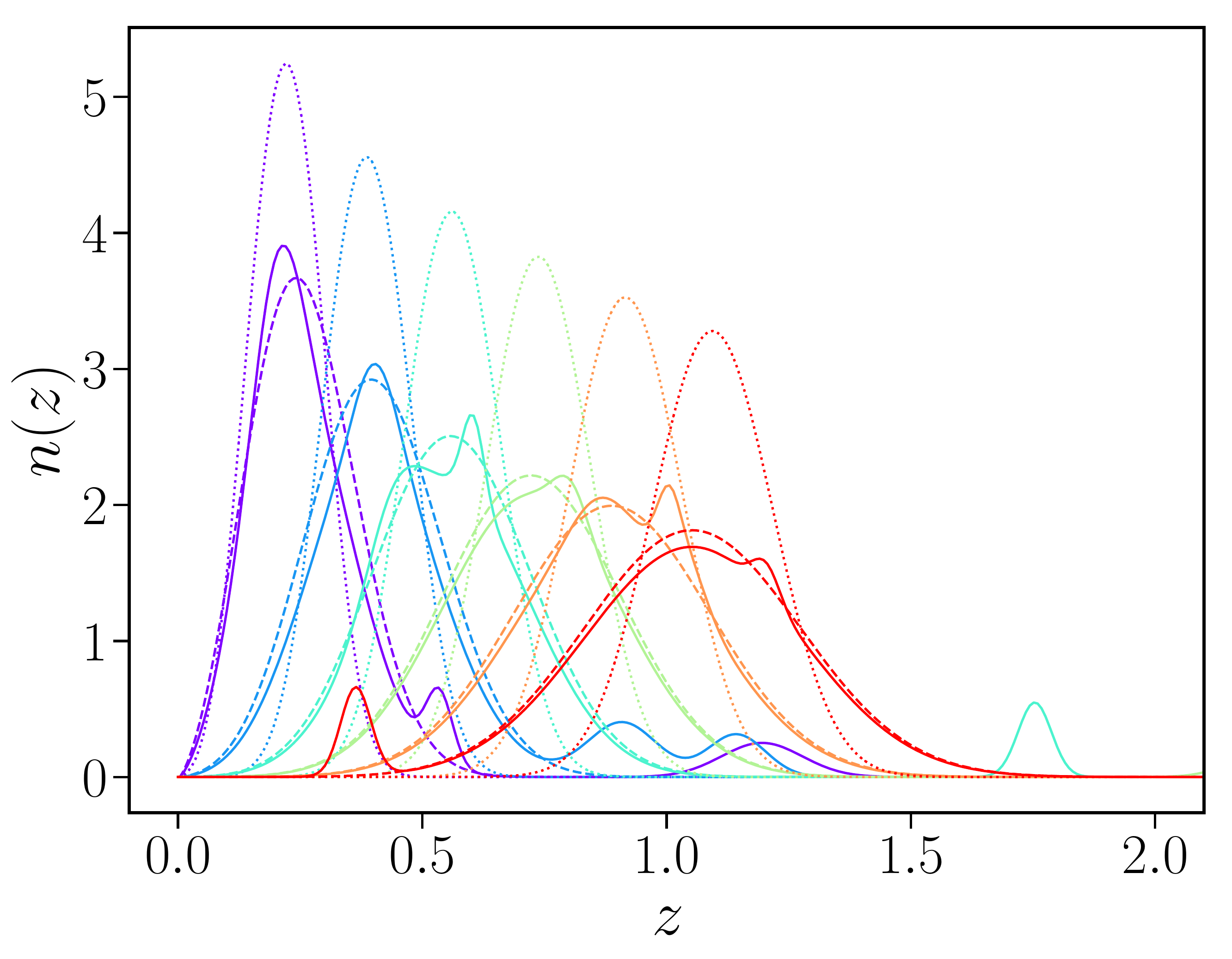}
\caption{Two of the four different $n(z)$ adopted in our comparison. The dotted curves refer to the case of Gaussian photometric distributions with $\sigma_z = 0.05(1+z)$ as discussed in the text (see Sect.~\ref{subsec:impact_on_lensing}), the dashed curves to broader Gaussians ($\sigma_z = 0.1(1+z)$), while the solid lines are the $n(z)$ built from the broader Gaussians with the inclusion of `catastrophic outliers' and more pronounced peaks in the distributions.}
\label{fig:nz_toy_model}
\end{figure}

\begin{figure*}
\centering
\includegraphics[width=\textwidth]{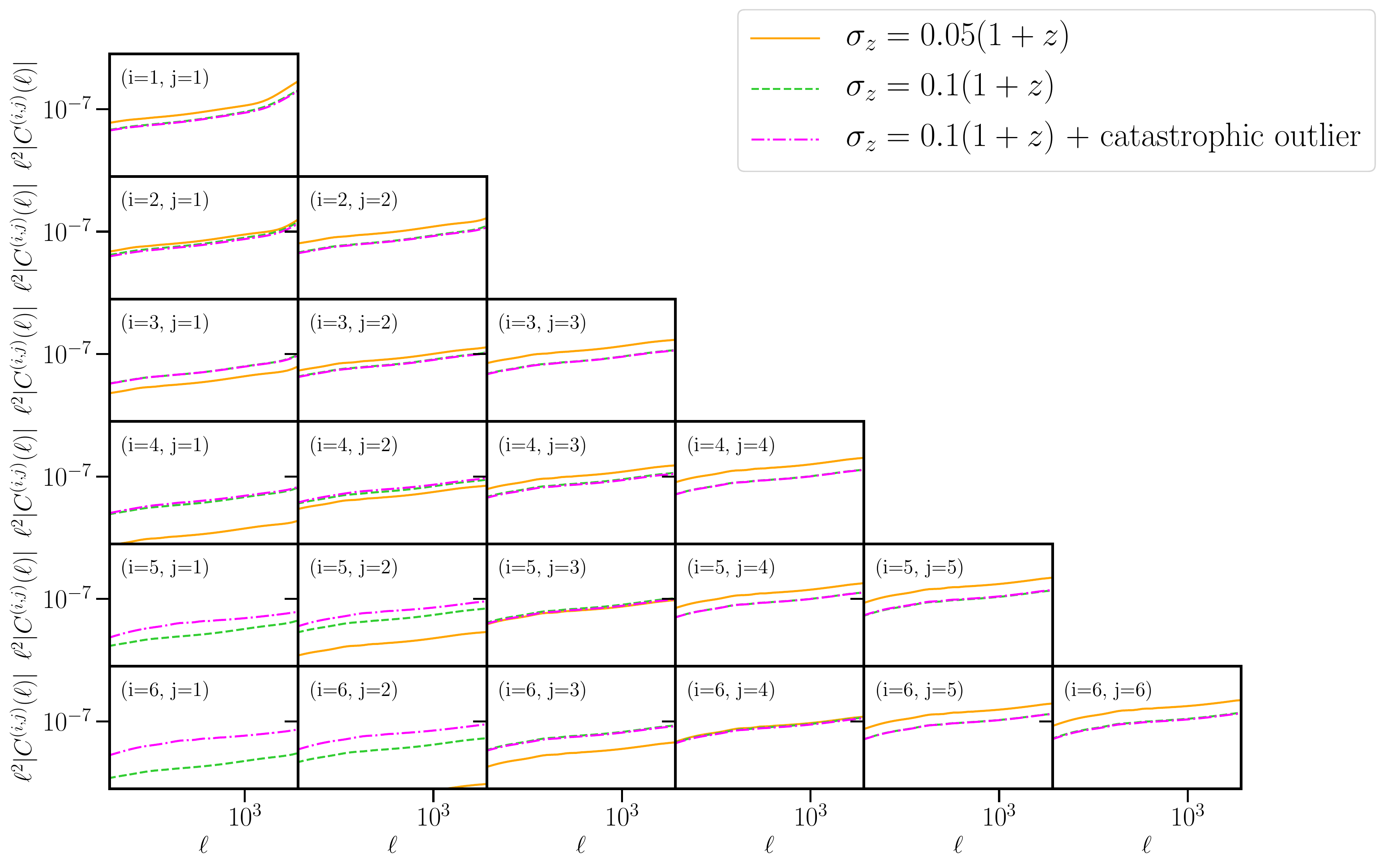}
\caption{The impact of the $n(z)$ on the projected II angular power spectrum. The IA is generated assuming the same setup as \ref{tab:halo_model_setup}, while the $n(z)$ are respectively: the fiducial Gaussian $n(z)$ adopted in the rest of the paper (solid orange lines), broader Gaussian distributions (dashed green lines) and borader Gaussians with the superpositions of `catastrophic outliers' and peaks (dash-dotted magenta lines).} 
\label{fig:cl_gaussians_nz_impact_II}
\end{figure*}

We illustrate here the impact on the choice of $n(z)$ in predicting the IA signal. As discussed in Sect.~\ref{sec:results}, the specific choice of the $n(z)$ distribution plays an important role in enhancing the II term, changing the balance between the different IA components. Here, we try to disentangle which feature has the largest impact on modulating the magnitude of the II term and its scale dependence. For this exercise, we use the IA model referred as (i), where the IA dependence on the red central galaxy luminosities is modelled as a single power law. We generate three different distributions (\ref{fig:nz_toy_model}), which progressively include a new feature. Fig. \ref{fig:cl_gaussians_nz_impact_II} illustrates our findings. We start with the Gaussian distributions adopted in the paper and presented in Sect.~\ref{subsec:impact_on_lensing} (solid orange lines), then we broaden them by increasing the standard deviation per bin, $\sigma_z^\mathrm{broad} = 0.1(1+z)$ (dashed green lines): this increases the amplitude of the II power spectra in the off-diagonal terms, due to the overlap of the tails of the distributions from different adjacent bins; on the diagonal terms, the broadening slightly reduces the II power spectrum. We then introduce `catastrophic outliers', which we generate as Gaussian islands centred on random points extracted from the original $dN/dz$, with a similar approach as \citet{Samuroff2019}. The presence of the outliers increases the II contribution (dash-dotted magenta lines): this is particular prominent at low redshift, for highly separated $z-$bins, where the outliers introduce correlated pairs between bins that would otherwise been uncorrelated.


\bsp	
\label{lastpage}
\end{document}